\documentclass[useAMS]{mn2e}
\usepackage{graphicx}
\usepackage{epsfig}
\usepackage{amssymb}
\usepackage{lscape}
\usepackage{ulem}
\usepackage{txfonts}

\def\kms{km~s$^{-1}$}
\def\cm{cm$^{-2}$}
\def\lya{Ly$\alpha$}
\def\nhi{$N$(H\,{\sc i})}
\def\hi{H\,{\sc i}}
\def\SiII{Si\,{\sc ii}}
\def\MgI{Mg\,{\sc i}}
\def\MgII{Mg\,{\sc ii}}
\def\FeII{Fe\,{\sc ii}}
\def\TiII{Ti\,{\sc ii}}

\def\ZnII{Zn\,{\sc ii}}
\def\CrII{Cr\,{\sc ii}}
\def\CI{C\,{\sc i}}
\def\c2s{C\,{\sc ii}$^{\star}$}
\def\hkpc{$h_{70}^{-1}$ kpc}
\def\ts{T$_s$}

\def\hii{H\,{\sc i}~21\,cm}

\title[DLAs in the redshift desert] {HI content, metallicities and
spin temperatures of damped and sub-damped Lyman alpha systems in the 
redshift desert
($0.6 < z_{\rm abs} < 1.7$).\thanks{Based on observations made with ESO
Telescopes at the La Silla Paranal Observatory under programme IDs
082.A-0569, 166.A-0106, and 085.A-0258, and observations made with the
NASA/ESA Hubble Space Telescope, obtained from the data archive at the
Space Telescope Institute. STScI is operated by the association of
Universities for Research in Astronomy, Inc. under the NASA contract
NAS 5-26555. These observations are associated with programs 9051 and
12214.} }

\author[Ellison et al.] {Sara L. Ellison$^1$, Nissim Kanekar$^2$, J.
Xavier Prochaska$^3$, Emmanuel Momjian$^4$, 
\newauthor Gabor Worseck$^3$. \\
$^1$ Department of Physics and Astronomy, University of Victoria, Victoria, 
British Columbia, V8P 1A1, Canada.\\
$^2$ National Centre for Radio Astrophysics, TIFR, Ganeshkhind, Pune-411007, 
India.\\
$^3$ Department of Astronomy and Astrophysics, UCO/Lick Observatory, 
University of California, 1156 High Street, Santa Cruz, CA 95064, USA.\\
$^4$ National Radio Astronomy Observatory, 1003 Lopezville Rd, Socorro, NM 87801, USA. \\
}

\begin{document}

\maketitle

\begin{abstract}

The \hii\ absorption optical depth and the \nhi\ column density
derived from \lya\ absorption can be combined to yield the spin
temperature (\ts) of intervening damped Ly$\alpha$ absorbers (DLAs).
Although spin temperature measurements exist for samples of DLAs with
$z_{\rm abs} <0.6$ and $z_{\rm abs} >1.7$, the intermediate redshift
regime currently contains only two \hii\ detections, leading to a
``redshift desert'' that spans 4 gigayears of cosmic time. To connect
the low and high redshift regimes, we present observations of the
\lya\ line of six $0.6 < z_{\rm abs} < 1.7$ \hii\ absorbers.  The dataset is
complemented by both low-frequency Very Long Baseline Array (VLBA)
observations (to derive the absorber covering factor, $f$), and
optical echelle spectra from which metal abundances are
determined. Our dataset therefore not only offers the largest
statistical study of \hii\ absorbers to date, and bridges the redshift
desert, but is also the first to use a fully $f$-corrected dataset to
look for metallicity-based trends.

The metallicities of \hii\ absorbers are consistent with those of
optically-selected DLAs at the same redshift.  In agreement with
trends found in Galactic sightlines, we find that the lowest column
density absorbers tend to be dominated by warm gas. In the DLA regime,
spin temperatures show a wider range of values than Galactic data, as
may be expected in a heterogenous galactic population.  However, we
find that low metallicity DLAs are dominated by small cold gas
fractions and only absorbers with relatively high metallicities
exhibit significant fractions of cold gas.  Using a compilation of
\hii\ absorbers which are selected to have covering-factor corrected
spin temperatures, we confirm an anti-correlation between metallicity
and \ts\ at $3.4\sigma$ significance.  Finally, one of the DLAs in our
sample is a newly-discovered \hii\ absorber (at $z_{\rm abs} = 0.602$
towards J1431+3952), which we find to have the lowest $f$-corrected
spin temperature yet reported in the literature: \ts~$= 90 \pm 23$~K.
The observed distribution of \ts\ and metallicities in DLAs and the
implications for understanding the characteristics of the interstellar
medium in high redshift galaxies are discussed.

\end{abstract}


\section{Introduction}

The interstellar medium (ISM) gas in galaxies represents the
raw material for star formation through cosmic time.  With current
instrumentation, the most effective way to study this gas (in its
neutral, ionised and molecular phases) is with absorption spectra
towards bright, high redshift sources such as quasars and gamma-ray bursts 
(e.g. Rauch 1998; Vreeswijk et al. 2004; Wolfe, Gawiser \& Prochaska 2005;
Prochaska et al. 2007b; Ledoux et al. 2009).  The damped Lyman-$\alpha$ 
systems (DLAs) and sub-DLAs (with log~\nhi~$\ge$~20.3 and 
19.0~$<$~log~\nhi~$<$~20.3, respectively) have garnered particular
attention as the galactic scale absorbers that contain the bulk of the
neutral gas, and which can be readily identified in relatively low
resolution spectra (e.g. Wolfe et al. 1986; Storrie-Lombardi \& Wolfe
2000; Ellison et al. 2001; Peroux et al. 2003; Rao, Turnshek \& Nestor
2006).  There are now some 1000 DLAs identified, with the largest haul
coming from the Sloan Digital Sky Survey (SDSS; Noterdaeme et
al. 2009; Prochaska \& Wolfe 2009).  High resolution spectroscopic
follow-up has yielded chemical abundances for over 200 DLAs and
sub-DLAs (e.g. Prochaska et al. 2007a; Akerman et al. 2005; 
Dessauges-Zavadsky, Ellison \& Murphy 2009; Meiring et al. 2009).  
These abundance measurements have
been brought to bear on several issues of nucleosynthetic origin and
enrichment mechanisms (e.g. Pettini et al. 2000, 2002; Ellison, Ryan
\& Prochaska 2001; Prochaska et al. 2002; Dessauges-Zavadsky et
al. 2002; Ledoux, Bergeron \& Petitjean 2002; Cooke et al. 2011).

Despite this wealth of spectroscopic data, surprisingly little is
known about the optical emission properties (such as morphologies and
luminosities) of DLA and sub-DLA galaxies.  Whilst there has been some
success in imaging the host galaxies at low redshifts (e.g. Le Brun
et al. 1997; Chen \& Lanzetta 2003; Chun et
al. 2010;  Rao et al. 2003, 2011), only a handful of high $z$ galaxy
counterparts have been identified (e.g. Warren et al. 2001;
Weatherley et al. 2005;  Fynbo et al. 
2010, 2011; Noterdaeme et al. 2012). To address the difficulty of detecting 
high redshift DLA galaxies,
several novel observing techniques have been initiated, such as imaging below
the Lyman limit of a higher redshift absorber or using integral field
units (O'Meara, Chen \& Kaplan 2006; Fumagalli et al. 2010; Peroux et
al. 2011).  At least at lower redshifts, where the sample sizes are
sufficient to draw general conclusions, galaxies selected based on 
their \hi\ properties alone seem to represent a wide variety of galaxy 
morphologies and luminosities (e.g. Rao et al. 2003, 2011).

Ultimately, however, it is perhaps the physical conditions within the
ISM that may reveal more about the nature and processes in high
redshift galaxies.  There are a few promising avenues for making such
measurements in DLAs.  For example, in cases where molecular hydrogen
is detected (e.g. Ge \& Bechtold 1997; Ledoux et al. 2003;
Noterdaeme et al. 2008) the relative populations of
the rotationally excited states can be used to constrain temperatures
and densities in the ISM (e.g. Srianand et al. 2005; Noterdaeme et
al. 2007, 2010).  In cases where the background QSO is radio-loud, an
alternative measure of the temperature can be obtained through 
observations of the \hii\ line in absorption.  In combination with the
\hi\ column density (measured from the \lya\ absorption profile), the \hii\
optical depth yields the so-called spin temperature (\ts) of the neutral 
gas.  For the general case of multiple phases of neutral gas along 
the line of sight (e.g. a cold phase with a kinetic temperature of 
$\approx 40-200$~K and a warm phase with a kinetic temperature of $\approx 
5000 - 8000$~K; Wolfire et al. 2003), the derived spin temperature 
gives the column density weighted harmonic mean of the spin temperatures 
of the multiple phases, and thus the distribution of neutral gas in different 
phases of the ISM (e.g. Kanekar \& Briggs 2004). Since the detection of
molecular hydrogen in DLAs is biased towards the highest metallicity
systems (Petitjean et al. 2006; Noterdaeme et al.  2008), the use of
spin temperatures is an appealing option for a more general overview
of conditions in the neutral ISM.

Despite its promise, the collection of spin temperature measurements
in DLAs has been fairly slow, with the first 25 years of \hii\
absorption studies yielding only 10 detections in DLAs at all
redshifts, with only 3 detections at $z > 1.7$ (e.g. Brown \&
Roberts 1973; Wolfe \& Davis 1979; Wolfe, Briggs \& Jauncey 1981;
Wolfe et al. 1985; Lane et al. 1998; Chengalur \& Kanekar 1999). The
primary reason for the lack of spin temperatures is the paucity of
DLAs at suitable redshifts in front of radio-loud QSOs.  Moreover, the
relatively narrow bandwidth of spectrometers has generally made blind
\hii\ absorption surveys impractical.  The last decade has seen
notable improvement in \hii\ absorption statistics, due to progress on
several fronts.  First, optical spectroscopic surveys that include
large numbers of QSOs are now available to cross-correlate with
all-sky radio surveys. This approach has led to the detection of \hii\
absorption associated with DLAs and \MgII\ absorbers towards SDSS QSOs
(e.g. Kanekar et al. 2009b; Gupta et al. 2009; Srianand et al.  2010, 2012).
A more targeted approach is to design an optical search for DLAs
towards radio-loud QSOs (Ellison et al. 2001; Jorgenson et al. 2006;
Ellison et al. 2008).  To fully exploit these new surveys, there has
also been progress on the technical side, with improvements in the
frequency range accessible by large radio telescopes such as the
Westerbork Synthesis Radio Telescope (WSRT), the Green Bank Telescope
(GBT) and the Giant Metrewave Radio Telescope (GMRT). As a result of
these advances, the opportunity now exists to gather significant
numbers of \ts\ measurements in DLAs at high redshifts, $z \gtrsim 2$.

Almost immediately, the measurements of \ts\ in high redshift DLAs
yielded a surprise result: whereas $z_{\rm abs} < 0.5$ DLAs exhibit a
range of spin temperatures, the sample of absorbers at $z_{\rm abs} >
2$ appeared to be dominated by spin temperatures in excess of 1000 K
(Carilli et al. 1996; Chengalur \& Kanekar 2000; Kanekar \& Chengalur
2003; Kanekar et al. 2007; Kanekar et al. 2012).  Although York et al.
(2007) discovered the first low \ts\ DLA at $z_{\rm abs}>2$, whether or not
there is a wholesale evolution in \ts\ with redshift has remained a
hotly debated topic (e.g. Kanekar \& Chengalur 2003; Wolfe, Gawiser \&
Prochaska 2003; Curran et al. 2005; Kanekar et al. 2009c).  In
addition to being limited by statistics, there has been considerable
focus on the possible effect of the so-called covering factor, $f$
(e.g. Wolfe, Gawiser \& Prochaska 2003; Curran et al. 2005; Kanekar et
al. 2009a).  If the background radio source is extended, then the
foreground absorber will not fully cover it; this would imply a low
covering factor, $f<<1$. Since most \hii\ absorption studies actually
determine the ratio \ts/$f$ (and not \ts\ alone), assuming $f=1$ would
result in an over-estimate of the spin temperature.  It has also been
claimed (Curran \& Webb 2006) that due to cosmological geometry,
different covering factors are to be expected in different redshift
regimes.  In particular, absorbers at high redshifts, $z_{\rm abs} >
1.6$, are always at larger angular diameter distances than the
background QSOs, potentially leading to systematically lower covering
factors for higher redshift DLAs.  However, in a Very Long Baseline
Array (VLBA) study of radio-loud QSOs with foreground DLAs, Kanekar et
al. (2009a) found none of the 24 DLAs of their sample have $f < 0.4$,
and that there was no difference in the distributions of the covering
factors of the high-$z$ and low-$z$ DLA samples. Kanekar et
al. (2009a) hence concluded that low covering factors in high-$z$ DLAs
were not the cause of the apparent redshift evolution in the spin
temperature.

A second intriguing suggestion involving spin temperatures is the
possible anti-correlation between \ts\ and metallicity (Kanekar
et al. 2009c).  Since metals are the primary coolant in the 
ISM, one would expect the fraction of cold gas to increase as the
metallicity increases (Kanekar \& Chengalur 2001). Kanekar et al. 
(2009c) reported a tentative detection of this anti-correlation at 
$3.6 \sigma$ significance from a sample of 26 DLAs, of which 10 had 
measurements of (rather than limits on) both \ts\ and metallicity.  
Again, the absorber covering factors were available for only a subset 
 of the DLAs.

The main challenge in confirming the redshift evolution of \ts\ and 
the anti-correlation of \ts\ and metallicity is simple statistics.
For example, there remains a glaring
gap in the redshift coverage of \ts\ measurements between $0.6 <
z_{\rm abs} < 1.7$, the so-called \hii\ redshift desert.  There are
only two published \hii\ detections in DLAs in this regime, which
spans 4 gigayears in cosmic time (Brown \& Roberts 1973; Kanekar et
al. 2009b).  To fill this regime, several surveys have searched for
\hii\ absorption in strong \MgII\ systems that have been identified
from ground-based optical surveys (e.g.  Ellison et al. 2004;
Prochter, Prochaska \& Burles 2006).  Although these \hii\ surveys
have now yielded a significant number of \hii\ absorption detections in the
redshift desert (Gupta et al. 2007, 2009; Srianand et al.  2008;
Kanekar et al. 2009b), the critical \nhi\ measurements required to determine
the spin temperatures have been missing.  A second obstacle has been
determining covering factors for the full redshift range of \hii\
absorbers, since these corrections are required to confidently test
any trends in \ts.

In this paper, we present a comprehensive set of multi-wavelength data
for six \hii\ absorbers in the redshift desert.  This is an effective
way to improve spin temperature statistics since the main limitation
of identifying systems with \hii\ detections has already been
addressed.  The data span six decades in rest wavelength ranging from
near ultra-violet (UV) absorption at 1216 \AA\ through the optical and
out to the redshifted 21~cm line frequency in the radio regime.  The
sample is selected primarily based on \hii\ absorption in a sample of
strong \MgII\ absorbers (Gupta et al. 2009; Kanekar et al. 2009b).
The near-UV data cover the \lya\ transition, from which the \hi\
column density of the absorber can be measured.  We also present VLBA
images of all of the QSOs in order to determine their low-frequency
covering factors.  Finally, we complement the dataset with optical
echelle spectra that permit the measurement of the column densities of
a variety of elements.  Combined with the \nhi\ measurements from the
\lya\ data, we can hence also determine metal abundances.  The dataset
presented here is therefore unique at this redshift and represents the
first systematic study of metallicities and spin temperatures in the
redshift desert.  In this paper, we focus on the metallicities of
\hii\ absorbers.  In a companion paper (Kanekar et al. in preparation)
we will address the issue of \ts\ evolution.

\section{Observations}

\begin{center}
\begin{table}
\begin{tabular}{lccccc}
\hline 
QSO  & $z_{\rm em}$ & $z_{\rm abs}$ &  NUV & Optical  & $S_{\rm 1.4~GHz}$ \\
     &              &               &  mag.& mag.     & (Jy) \\  
\hline 
B0105$-$008 & 1.374 & 1.37078 & 18.8 &  $g=17.7$ & 0.93  \\
B0237$-$233 & 2.223 & 1.67235 & 19.5 &  $B=16.8$  & 6.26  \\
B0801+303   & 1.451 & 1.19110 & 19.8 &  $g=18.3$ & 1.27  \\
J1431+3952  & 1.215 & 0.60190 & ...  &  $g=16.5$ & 0.22  \\
J1623+0718  & 1.648 & 1.33567 & ...  &  $g=17.7$ & 0.081 \\
B2355$-$106 & 1.639 & 1.17303 & 20.7 &  $g=18.8$ & 0.77  \\
\hline
\end{tabular}
\caption{\label{sample_tab} The sample of QSOs with \hii\ absorption
studied in this work.  NUV magnitudes are taken from the GALEX survey,
while the 1.4~GHz flux densities are from the National Radio Astronomy
Observatory (NRAO) Very Large Array (VLA) Sky Survey (NVSS; Condon et al. 
1998). All DLA absorption redshifts quoted in this paper are values derived 
from the peak \hii\ absorption (Kanekar et al. 2009b; Gupta et al. 2009; 
see Table \ref{21cm_tab}). }
\end{table}
\end{center}

\subsection{\hii\ spectroscopy with the GBT}\label{GBT_sec}

\begin{figure}
\centerline{{\resizebox{3.3in}{!}
{\includegraphics{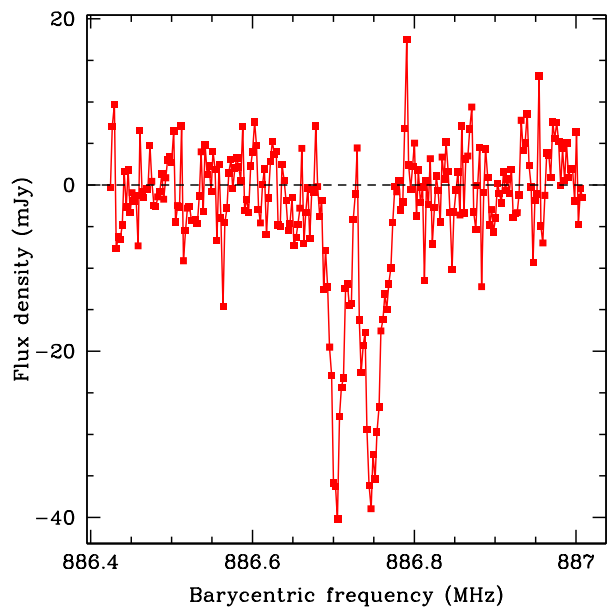}}}}
\caption{\label{fig:J1431_GBT} GBT spectrum of the \hii\ absorption at 
$z_{\rm abs} = 0.602$ towards J1431+3952. }
\label{fig:1431}
\end{figure}

\begin{figure*}
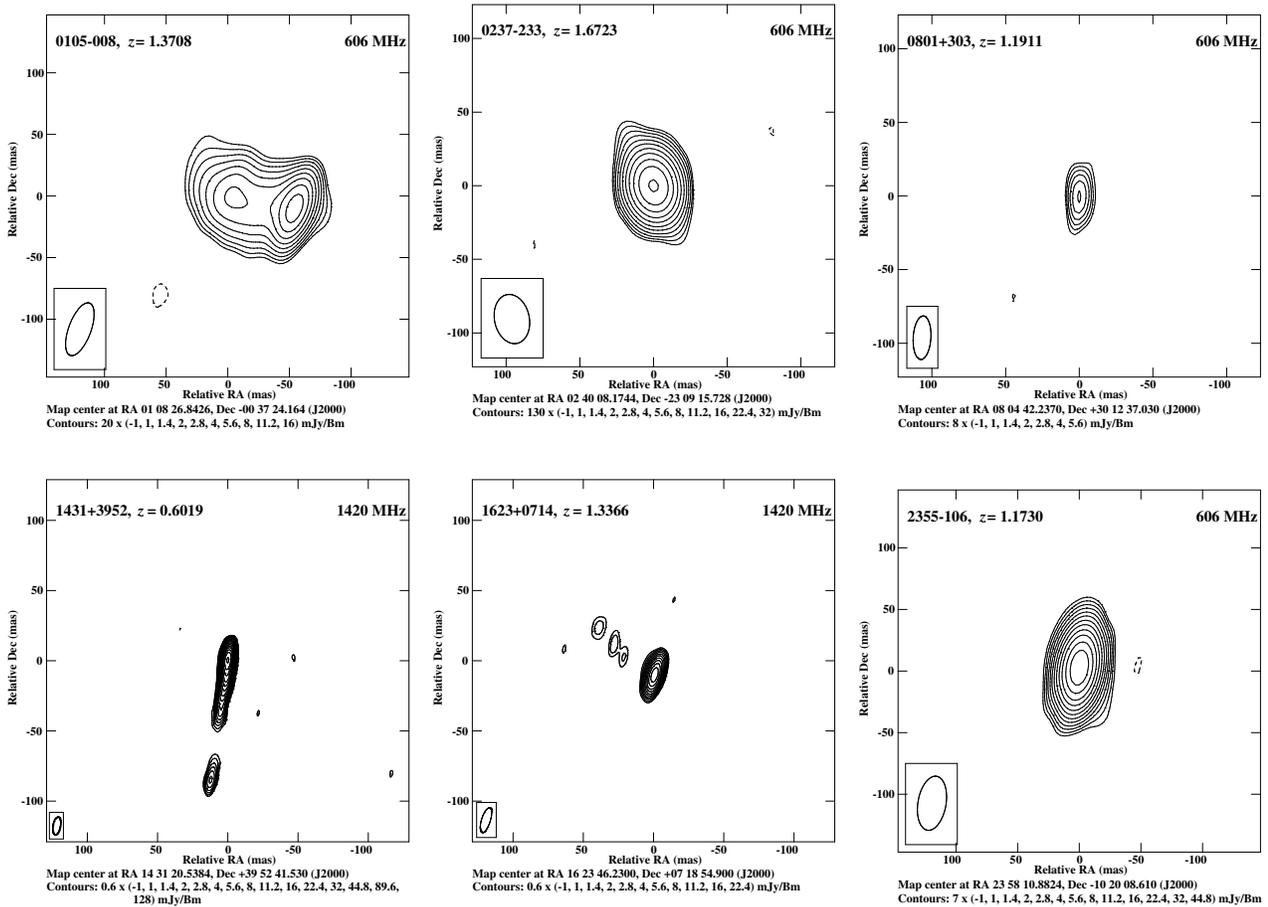

\begin{center}
\epsfig{file=fig2a.eps,width=2.2in}
\epsfig{file=fig2b.eps,width=2.2in}
\epsfig{file=fig2c.eps,width=2.2in}
\epsfig{file=fig2d.eps,width=2.2in}
\epsfig{file=fig2e.eps,width=2.2in}
\epsfig{file=fig2f.eps,width=2.2in}
\end{center}
\caption{VLBA images of the 6~quasars in the sample, ordered (from top
left) by quasar right ascension. The map frequency is at the top right of each
panel.}
\label{fig:vlba}
\end{figure*}

The targets in our sample are listed in Table \ref{sample_tab}.  Of
the 6 absorbers included in the current sample, 5 have previously
published \hii\ detections with the GMRT or the GBT (Kanekar et
al. 2009b; Gupta et al. 2009).  The final target, J1431+3952, is a new
\hii\ absorption detection that we present here for the first
time. For completeness, the details of the \hii\ detections of all six
absorbers in our sample, including J1431+3952, are given in Table
\ref{21cm_tab}.

We initially observed J1431+3952 with the GBT on  September 29 2009,
using the PF1-800 receiver, the Spectral Processor and a 2.5~MHz
bandwidth centred at 886.645~MHz and sub-divided into 1024
channels. This was part of a large new GBT \hii\ absorption survey of
strong \MgII\ absorption systems (proposal AGBT10B\_008; PI
Kanekar). Position-switching with an On/Off time of 5 minutes was used
to calibrate the bandpass, while the system temperature was determined
via a blinking noise diode. The total on-source time was 15~minutes,
sub-divided into 10-second integrations.

The initial observations of J1431+3952 revealed a strong absorption
feature at the expected redshifted \hii\ line frequency. We hence
repeated the observations on October 20 2010 and January 21 2011,
using a 1.25~MHz bandwidth to obtain higher velocity resolution, as
well as longer integration times (30~minutes on-source at each epoch).
These observations confirmed the 21~cm absorption by a detection of the
doppler shift in the observed \hii\ line frequency, due to the Earth's
motion around the Sun.

The GBT data were analysed using the NRAO {\sc AIPS++} single-dish
package {\sc DISH}, following standard procedures. The data were first
visually inspected for radio frequency interference (RFI) and Spectral
Processor failures, with corrupted data edited out.  After editing,
the individual spectra were calibrated and averaged together, in the
barycentric frame, to measure the flux density of J1431+3952 at
886.64~MHz. A second-order spectral baseline was then fit to each
10-second spectrum, during calibration, and subtracted out, before
averaging the residual spectra together, again in the barycentric
frame, to produce the final spectrum.

We note that very different source flux densities were obtained in the
October and January observing runs ($179-240$~mJy), although the depth
of the absorption features agree within the noise. The strong
out-of-band RFI present above 887~MHz makes it difficult to accurately
calibrate the flux density scale in such total power measurements. We
have used the average flux density from the two runs ($210$~mJy) to
compute the \hii\ optical depth; the final quoted error includes the
above systematic error in the flux density scale calibration, assumed
to be $\pm 30$~mJy, i.e. $\approx 15$\%.

The final \hii\ spectrum from the higher-resolution observations of
2010 October and 2011 January is shown in Fig.~\ref{fig:1431}; this
has a velocity resolution of 0.83~\kms\ (after Hanning-smoothing and
re-sampling) and a root mean square (RMS) noise of 4.5~mJy per
independent 0.83~\kms\ spectral channel. Two clear features are
visible in the spectrum, with depths of $\sim 40$~mJy; each has high
statistical significance and was detected on all three observing
runs. The integrated \hii\ optical depth is $(3.07 \pm 0.34)$~\kms,
where we have added in quadrature the errors on the integrated flux
density and the flux density scale.

\begin{center}
\begin{table*}
\begin{tabular}{lcccccc}
\hline 
QSO  & Telescope  & Resolution &  On-source Time   &  RMS  & $\int \tau$ dV & Reference\\
     &            & (\kms)     & (hours)           & (mJy) & (\kms)         &          \\
\hline 
B0105$-$008   &  GMRT  & 2.0 & 3.5 & 3.3 & $0.995 \pm 0.023$ & Kanekar et al. (2009b)\\
B0237$-$233 & GBT & 0.69 & 1.5 & 56 & $0.076 \pm 0.016$ & Kanekar et al. (2009b)\\
B0801+303     &  GMRT  & 3.6 & 5.0 & 2.2 & $0.305 \pm 0.025$ & Kanekar et al. (2009b)\\ 
J1431+3952    &  GBT  &  0.8 &  1.0   & 8.3 &  $3.07 \pm 0.34$  & This work\\
J1623+0718    &  GMRT  & 3.9 &11.7 &1.0  &  $0.916 \pm 0.099$  & Gupta et al. (2009)\\
B2355$-$106   &  GMRT  & 1.8 & 4.0 & 2.5 & $0.256 \pm 0.062$ & Kanekar et al. (2009b)\\
\hline
\end{tabular}
\caption{\label{21cm_tab} Summary of \hii\ observations.}
\end{table*}
\end{center}

\subsection{High spatial resolution radio imaging with the VLBA}
\label{VLBA_sec}

\begin{table*}
\begin{center}
\begin{tabular}{lccccccccc}
\hline 
         QSO  & On-source &Frequency  & Beam      &  RMS     &  $S_{\rm core}$ & $S_{\rm int}$ & Ang. size & Spatial extent & $f$  \\
              & time (hours) & (MHz)	   & (mas$^2$) & (mJy/Bm) &      (Jy)      & (Jy)           & mas~$\times$~mas & pc$^2$ & \\
\hline 
B0105$-$008   & 2.0 & 606  & $45.3 \times 18.1$ & 4.0  & 0.403  & 1.26  & $12.2 \times 0.0$ & $104 \times 0$ & 0.32 \\
B0237$-$233   & 2.0 &  606  & $33.8 \times 23.6$ & 23   & 4.95  & 5.53  & $11.3 \times 9.7$ & $97 \times 83$ & 0.90 \\
B0801+303     & 2.0 &  606  & $29.6 \times 12.0$ & 2.3  & 0.053 & 2.45  & $12.1 \times 4.7$ & $101 \times 39$ & 0.02 \\
J1431+3952    & 1.6 &  1420 & $12.9 \times ~5.3$ & 0.14 & 0.065 & 0.207 & $3.2 \times 0.0$ & $21 \times 0$ & 0.32 \\
J1623+0718    & 2.0 &  1420 & $18.4 \times ~6.3$ & 0.21 & 0.028 & 0.081 & $4.3 \times 1.4$ & $36 \times 12$ & 0.34 \\
B2355$-$106   &  2.0 & 606  & $44.4 \times 22.8$ & 2.2  & 0.42  & 0.42  & $10.8 \times 2.2$ & $90 \times 18$ & 1.0  \\
\hline
\end{tabular}
\caption{\label{table:vlba} Summary of VLBA observations and results. 
The covering factor, $f$, is the ratio of the flux in the core
($S_{\rm core}$) to the total integrated flux ($S_{\rm int}$). A spatial extent
of zero indicates a size much smaller than the synthesised beam.}
\end{center}
\end{table*}

We used the VLBA in proposals BK159 and BK174 (PI Kanekar) to obtain
high spatial resolution images of five of the background quasars of
our sample, at frequencies close to the redshifted \hii\ line
frequency. The observations of four targets (B0105$-$008, B0237$-$233,
B0801+303 and B2355$-$106) used the 606~MHz receivers, while
J1431+3952 was observed with the 1.4~GHz receivers. The 606 MHz
observations were carried out on 11 September 2009 and 4 January 2010
using two adjacent 2 MHz baseband channel pairs, while the 1.4 GHz
observations were carried out on 27 December 2011 using four adjacent
16 MHz baseband channel pairs, both with right- and left-hand circular
polarizations, and sampled at 2 bits. The 606 MHz and the 1.4 GHz data
were correlated using the VLBA hardware correlator and the VLBA DiFX
software correlator (Deller et al. 2011) in Socorro, NM, respectively,
with a 4 second correlator integration time and 32 spectral channels per
respective baseband channel. The total bandwidth of the data was 4 and
64 MHz for the 606 MHz and 1.4 GHz observations, respectively. Phase
referencing was not used. Bandpass calibration was carried out using
observations of the strong calibrators 3C345, 3C454.3, 3C48 and
3C147. The total on-source time was $\sim 2$~hours for the 606~MHz
targets and 1.6~hours for J1431+3952. In addition, we also retrieved
and analysed an archival VLBA 1.4~GHz dataset on the last source in
our sample, J1623+0718 (proposal BG0189, PI Gupta); these observations had a
similar setup to the above noted 1.4 GHz observations but used four
adjacent 8 MHz baseband channel pairs, resulting in a total
bandwidth of 32 MHz.

The VLBA data were analysed in ``classic'' {\sc AIPS}, using standard
procedures. After applying the initial ionospheric corrections and
editing data affected by RFI, the flux density scale was calibrated
using the measured antenna gains and system temperatures. This was
followed by fringe-fitting for the delay rates and bandpass
calibration, after which the data towards each target were averaged in
frequency to produce a single-channel visibility dataset. The 606~MHz
data were found to be of relatively low quality, with only 6-7 working
antennas and poor ionospheric conditions. A number of cycles of
self-calibration and imaging were used to derive the antenna-based
gains for each target, with phase-only self-calibration used for the
606~MHz datasets and amplitude-and-phase self-calibration for the
1.4~GHz datasets. For each source, the above cycles were repeated
until no improvement was seen on further self-calibration. The final
images obtained from the above procedure are shown in
Fig.~\ref{fig:vlba}, in order of increasing right ascension. For each source,
the synthesised beam and off-source RMS noise are listed in Table
\ref{table:vlba}.

Finally, the {\sc AIPS} task {\sc JMFIT} was used to fit a gaussian
model to the radio continuum images, to measure the compact flux
density of each quasar. A single gaussian was used as the model for
all sources but J1431+3952 and B0105$-$008, where a 2-gaussian model
was used due to the clear presence of a jet or extended structure.
The VLBA results are summarised in Table~\ref{table:vlba}, whose
columns contain (1)~the source name, (2)~ the on-source time, (3)~the
observing frequency (MHz), (4)~the synthesised beam, (5)~the RMS noise
(mJy/Bm), (6)~the integrated core flux density (Jy) from the VLBA
images, obtained using {\sc JMFIT}, (7)~the total flux density from
low resolution interferometric images at a similar frequency (Jy),
(8)~the deconvolved angular size of the core radio emission, in
mas~$\times$~mas, (9)~the spatial extent of the core emission at the
absorber redshift, and (10)~the covering factor $f$, obtained by
taking the ratio of the core flux density to the total flux
density. We do not quote formal errors on the covering factor for two
reasons.  First, the VLBA and low-frequency measurements were not
carried out simultaneously, which means that it is hard to account for
source variability in the error estimates.  Second, it is possible
that the foreground DLA also covers some of the extended radio
emission, making the values of $f$ lower limits.  For two sources
(B0105$-$008 and J1431+3952), the deconvolved angular size and the
spatial extent are $0.0$ along one axis; this implies that the core is
unresolved along this axis, with an angular size significantly smaller
than the VLBA synthesised beam.

\subsection{Near UV spectroscopy with HST/COS}\label{HST_sec}

\begin{figure*}
\centerline{\rotatebox{270}{\resizebox{14cm}{!}
{\includegraphics{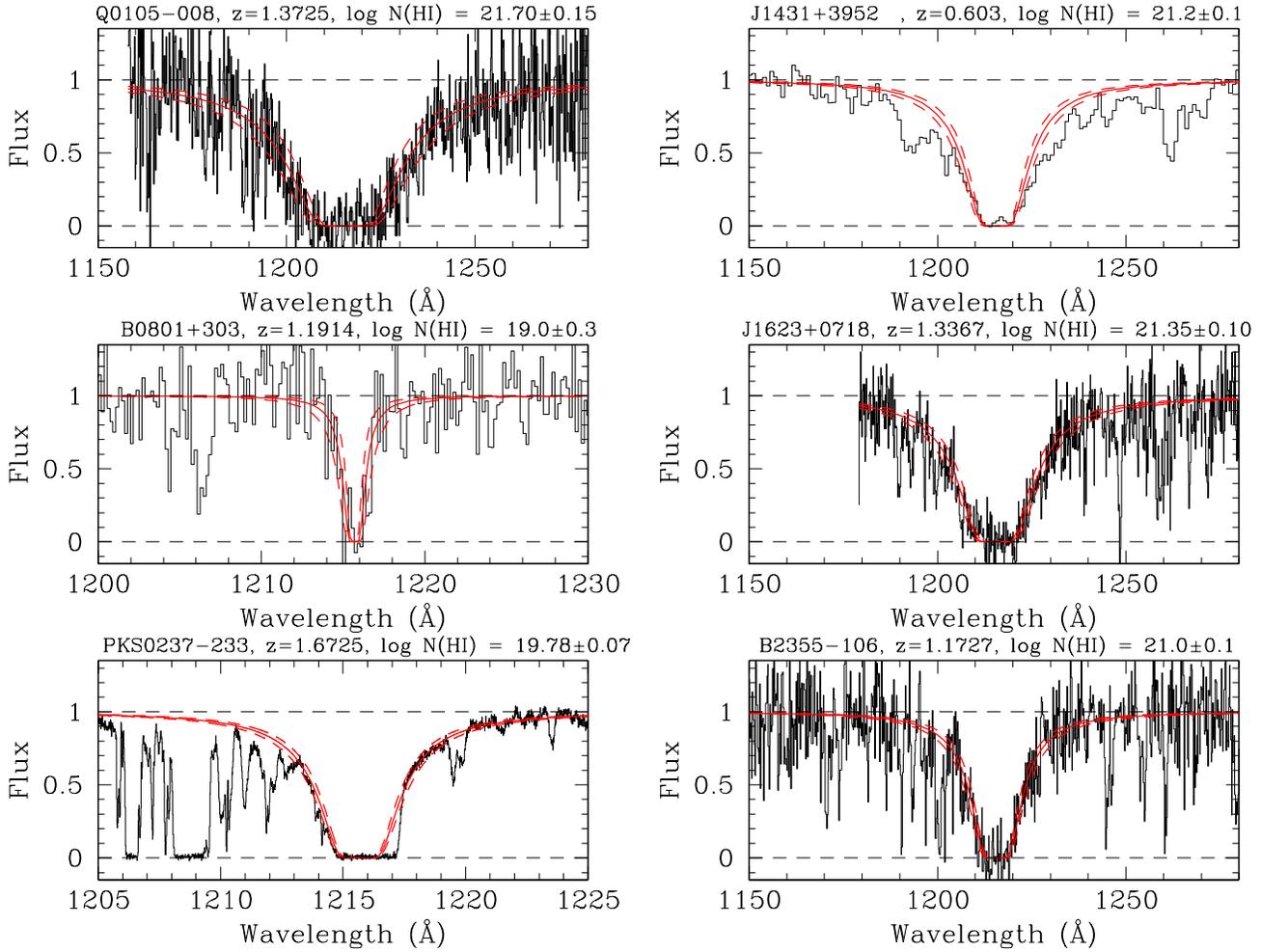}}}}
\caption{\label{dla_fits} Voigt profile models to the \lya\ line for
the absorbers in our sample are shown as red curves.  The dashed
lines indicate the uncertainties given in the panel captions.  }
\end{figure*}

Hubble Space Telescope (HST) Cosmic Origins Spectrograph (COS) spectra
have been obtained for 4 of the 6 absorbers in our sample (proposal ID
12214, PI Ellison).  For B0237$-233$, the \lya\ line is covered by a
Very Large Telescope (VLT) UV and Visual Echelle Spectrograph
(UVES) spectrum and discussed in Section \ref{echelle_sec}. The final
target, J1431+3952, has an existing archival HST Space Telescope
Imaging Spectrograph (STIS) spectrum from Cycle 10 (proposal ID 9051,
PI Becker) for which we use the pipeline-reduced spectrum available
from the archive.

The HST/COS data were obtained with
the near-UV (NUV) channel and the G230L grating which yields a full width at
half maximum (FWHM) resolution of $\sim$0.8 \AA.  The target was
acquired using Guide Star Catalog 2 (GSC2) coordinates in
{\sc ACQ/IMAGE} mode in order to minimise acquisition overheads.
Spectroscopic data were obtained in {\sc TIME-TAG} mode with at least two
exposures per target, each in a different {\sc FP-POS} position.  Total
exposure times and central wavelength settings are given in Table
\ref{HST_tab}.

\smallskip

The COS spectra were homogeneously reduced using CALCOS v2.15.4 and
custom routines for background estimation and co-addition of
individual exposures in the Poisson limit (e.g.\ Worseck et
al. 2011). The data were flat-fielded, deadtime-corrected,
doppler-corrected and wavelength-calibrated using the default {\sc CALCOS}
calibration files as of September 2011. For 1-d extraction we adjusted
the source extraction window to a rectangular box (height 17 pixels,
corresponding to 5$\times$ the spatial FWHM) to preserve integer
extracted counts in the Poisson regime, and to maximise the signal-to-noise 
ratio (S/N) for
our well-centred point sources. Likewise, the background extraction
windows were maximised (150 pixels each), and background smoothing by
the pipeline turned off to investigate
spectral variations of the background. After visual inspection, the
smoothed background was determined by a 51-pixel running average in
wavelength. Individual exposures were co-added by adding the integer
source counts and the non-integer mean background per pixel on the
homogeneous NUV wavelength grid, accounting for individual pixel
exposure times induced by offsets in dispersion direction. The
counts were then flux-calibrated with the NUV flux calibration curve and
the pixel exposure times.  The S/N ratios per pixel (in the continuum on either
side of the \lya\ absorption) of the final combined
spectra are given in Table \ref{HST_tab}.

\smallskip

The Starlink package {\sc DIPSO} was used to model a Voigt-profile
absorption to the normalised spectra.  To normalise the spectra, a
cubic spline was fit through unabsorbed regions of the QSO continuum
and the data then divided by the polynomial fit.  Despite the
relatively low S/N, the \nhi\ could usually be determined to within
$\pm$ 0.1 dex, which is typical for moderate resolution spectra of low
S/N (e.g. Russell, Ellison \& Benn 2006).  The dominant source of
error in \nhi\ is from uncertainty in the continuum placement.  This
is a particular issue for the proximate DLA towards B0105$-$008 where
the \lya\ absorption lies on top of the QSO's \lya\ emission.
Nonetheless, errors on \nhi\ are less $\le$ 0.15 dex for all the
\textit{bona-fide} DLAs (log \nhi\ $\ge$ 20.3) in our sample. Table
\ref{HST_tab} lists the measured \nhi\ column densities for both the
COS and STIS data and Figure \ref{dla_fits} presents the \lya\ fits to
the absorbers in our sample.

From Table \ref{HST_tab} it can be seen that 4/6 absorbers in our
sample qualify as DLAs; indeed these 4 have high \hi\ column densities, 
in excess of 10$^{21}$ \cm.  The absorbers towards B0801+303 and 
B0237$-$233 have \nhi\ values below the DLA threshold, but can still be well
fit with reasonable accuracy and fall into the category of sub-DLAs.

\begin{center}
\begin{table*}
\begin{tabular}{llccccl}
\hline 
QSO  & Telescope/ & Grating/    &         \#      & Time  &  S/N         &  log N(HI)\\
     & Instrument & wavelength (\AA)  &  orbits   & (s)   & pix$^{-1}$ & \\
\hline 
B0105$-$008  &  HST/COS & G230L/2950  & 1 & 2,454  & 4 & 21.70$\pm$0.15\\ 
B0237$-$233 & VLT/UVES & 346+580 & ... & 42,927 & 40 & 19.78$\pm$0.07 \\ 
B0801+303  & HST/COS & G230L/2635  & 2 & 5,619  & 5 & 19.0$\pm$0.3\\    
J1431+3952 & HST/STIS& G230L       & 1 & 2,292  & 15 & 21.2$\pm$0.1 \\   
J1623+0718 & HST/COS & G230L/2950  & 4 & 11,499 & 5 & 21.35$\pm$0.10\\  
B2355$-$106  & HST/COS & G230L/2635  & 4 & 11,451 & 4 & 21.0$\pm$0.1\\    
\hline
\end{tabular}
\caption{\label{HST_tab} Summary of \lya\ observations.  
The S/N is measured in the continuum on either side of the \lya\ absorption
and is derived from the RMS noise.  However, at low S/N (the Poission
limit) the formal S/N is about a factor of two lower.}
\end{table*}
\end{center}

\subsection{Optical echelle spectroscopy with Keck/HIRES and VLT/UVES}\label{echelle_sec}

High resolution echelle spectra have been obtained for B0105$-$008,
B1430$-$178, J1431+3952 and J1623+0718 using the High Resolution
Echelle Spectrograph (HIRES; Vogt et al. 1994) on the Keck telescope
and UVES on the VLT.  Our HIRES observations of B0105$-$008 were
supplemented with archival UVES data of this target (proposal
082.A-0569, PI Srianand) in order to extend the wavelength coverage of
the spectrum.  Archival UVES data were also downloaded and reduced for
B2355$-$106 (proposal 085.A-0258, PI Srianand).  The same UVES program
also obtained data for J1623+0718 which we use to supplement our HIRES
data (which extend further into the blue and cover the UVES inter-arm
and chip gaps).  B0237$-$233 was a target in the ESO Large Program
166.A-0106 (PI Bergeron) and its UVES data cover both the \lya\ and
metal transitions.  We do not have echelle spectra for the final
target in our sample (B0801+303).  Table \ref{hires_tab} summarises
the available observations, instrument set-ups, wavelength coverage
and final S/N ratios.

\begin{center}
\begin{table*}
\begin{tabular}{lcccccc}
\hline 
QSO  & Instrument &  Date/ &    Time   & Instrument & Wavelength &S/N           \\
     &            &  Archive ID &  (s)    & set-up     & coverage (nm) &pix$^{-1}$ \\
\hline 
B0105$-$008  &  HIRES  & Jan. 2008 & 7,200  & 0/1.250  & 307-593          & 10--20 \\
           &  UVES   & 082.A-0569 &14,790 & 390+580  & 350-450, 476-684 & 20--50  \\
B0237$-$233 & UVES & 166.A-0106 & 42,927 & 346+580 & 305-1008 &  40-150 \\
              &      &            & 48,030 & 437+860 & (combined) & \\
J1431+3952 & HIRES    & Jan. 2012 &  1,200  &  0/1.156   &  317-603   & 6--15 \\
J1623+0718 &  HIRES  & Sept. 2010 &  7,200 &  0/1.0005  & 320-588  & 10\\
          &  UVES  & 085.A-0258 &  13,356 &  390+580  &  349-451, 479-680  & 10--40\\
B2355$-$106  &  UVES & 085.A-0258  & 23,373  & 390+580   &    350-450, 476-684 & 8--25 \\
\hline
\end{tabular}
\caption{\label{hires_tab} Summary of our optical echelle
observations.  The quoted S/N range is over the full spectral coverage. 
For spectra obtained directly by us, we provide the date of the observations. 
For UVES archival data, the proposal ID is given. The instrument set-up for 
HIRES refers to the echelle/cross disperser angles.}
\end{table*}
\end{center}

UVES is a dual-arm echelle spectrograph with a grating cross
disperser (Dekker et al. 2000).    The data presented in
this paper were obtained in service mode (or from the archive) using 
various combinations of standard wavelength settings.  The
actual wavelength coverage is given in Table \ref{hires_tab} as
measured directly from the spectra.  The CCDs were binned 2x2 and a 1
arcsecond slit was used, resulting in a typical resolution of
$R=42,000$.  The spectra were reduced using a custom version of the
Midas reduction pipeline, with manual intervention at several stages.
In particular, the tolerance of the wavelength solution was adjusted
to optimise the number of identified arc lines and accuracy of the
wavelength solution.  The typical accuracy of the wavelength solution
was 60--70 m/s.  The extracted spectra from each exposure were
converted to a common vacuum-heliocentric wavelength scale.  The
heliocentric correction was calculated externally, as the UVES fits
headers do not always accurately record this parameter.  Where
multiple exposures in a given setting were obtained, or where there is
spectral overlap between settings, the data were combined, weighting
by the inverse of the flux variances. The final extracted
1-dimensional spectrum was normalised by fitting the continuum with
Starlink's {\sc DIPSO} package.

HIRES data for three targets were obtained in PI mode.  B0105$-$008
was observed with HIRES on 12 January, 2008 with
the blue cross-disperser and collimator (i.e. HIRESb).  We employed
the C1 decker for a nominal spectral resolution of FWHM~$\sim 6$
\kms.  Two exposures totalling 7200s were acquired under
fair conditions.  J1431+3952 was observed on 16 January 2012 in good
conditions, also with the C1 decker and HIRESb, with a single exposure
of 1200s.  The HIRES data were reduced with the HIRedux
pipeline that is available within the XIDL software
package\footnote{http://www.ucolick.org/$\sim$xavier/IDL}.  The main
difference in the processing of the UVES and HIRES data occurs in the
normalization procedure.  Whereas normalisation is the final step in the UVES
reductions, the HIRedux package fits the continuum order-by-order
before conversion to a 1-dimensional spectrum.  The HIRES spectra are
also corrected to a vacuum heliocentric scale.

Mosaics of many of the metal line transitions detected in the HIRES
and UVES data are presented in Figures \ref{UM305_metals} to
\ref{2355_metals}.

\begin{center}
\begin{table*}
\begin{tabular}{lcccccccc}
\hline 
QSO  & $z_{\rm abs}$ & N(FeII) & N(SiII) & N(ZnII) & N(CrII) & N(MnII) & N(TiII) & N(NiII)\\
\hline 
B0105$-$008  & 1.37078 & 15.59$\pm$0.03 & $\ge$ 15.85 & 12.93$\pm$0.04 & 13.92$\pm$0.03 & 13.29$\pm$0.04 & 13.0$\pm$0.2& 14.25$\pm$0.04\\
B0237$-$233 & 1.67325 & 14.57$\pm$0.02 & 14.92$\pm$0.02 & 11.84$\pm$0.09 &12.72$\pm$0.06 & 12.24$\pm$0.06 & $<$11.46 & 13.51$\pm$0.04\\
J1431+3952 & 0.60190 &15.15$\pm$0.11 &... & 13.03$\pm$0.19&13.53$\pm$0.18 &13.07$\pm$0.14 &12.70$\pm$0.15 &...\\
J1623+0718 & 1.33567 &15.28$\pm$0.05 &15.78$\pm$0.05 &12.91$\pm$0.09 &13.55$\pm$0.06 & 13.06$\pm$0.06 & $<$12.54 & 14.11$\pm$0.07\\
B2355$-$106  & 1.17303 &15.08$\pm$0.10 &15.42$\pm$0.13 &12.76$\pm$0.17 & $<$13.2 &12.91$\pm$0.10 & $<$12.9 & 14.17$\pm$0.12\\
\hline
\end{tabular}
\caption{\label{coldens_tab} Column densities derived from the apparent optical depth method. No high resolution spectrum is available for B0801+303.  For non-detections, we quote 3$\sigma$ upper limits. Lower limits are derived from lines exhibiting some degree of saturation. }
\end{table*}
\end{center}

\begin{center}
\begin{table*}
\begin{tabular}{lcccccccc}
\hline 
QSO  & $z_{\rm abs}$ & [Fe/H] & [Si/H] & [Zn/H] & [Cr/H] & [Mn/H] & [Ti/H] & [Ni/H]\\
\hline 
B0105$-$008  & 1.37078 & $-1.56$ & $\ge-1.36$ & $-1.40$ & $-1.42$ & $-1.89$ & $-1.61$ & $-1.65$\\
B0237$-$233 & 1.67325 & $-$0.66 & $-$0.37 & $-$0.57 & $-$0.70 & $-$1.02 & $<-$1.23 & $-$0.47 \\
J1431+3952 & 0.60190 &$-$1.50 &... &$-$0.80 &$-$1.31 &$-$1.61 &$-$1.41 &...\\
J1623+0718 & 1.33567 & $-$1.52 &$-1.08$ &$-1.07$ &$-1.44$ & $-1.77$ &$<-1.72$ & $-1.44$\\
B2355$-$106  & 1.17303 &$-$1.37 &$-$1.09 &$-$0.87 &$<-$1.44 &$-$1.57 &$<-$1.01 &$-$1.03\\
\hline
\end{tabular}
\caption{\label{abund_tab} Elemental abundances. No high resolution spectrum is 
available for B0801+303.}
\end{table*}
\end{center}

\begin{center}
\begin{table}
\begin{tabular}{lcccc}
\hline 
QSO  & $z_{\rm abs}$ & log N(HI) &  T$_s$ (K) & [Zn/H] \\
\hline 
B0105$-$008  & 1.37078 & 21.70$\pm$0.15 & 305$\pm$45 & $-$1.40\\
B0237$-$233 & 1.67235 & 19.78$\pm0.07$ & 390$\pm$127 & $-$0.57 \\
J1431+3952 & 0.60190 & 21.2$\pm$0.1   & 90$\pm$23 & $-0.80$\\
J1623+0718 & 1.33567 & 21.35$\pm$0.10 & 460$\pm$105 & $-$1.07\\
B2355$-$106  & 1.17303 & 21.0$\pm$0.1   & 2145$\pm$570 & $-$0.87\\
\hline
\end{tabular}
\caption{\label{summary_tab} Summary of DLA properties.  Spin temperature
values take into account the covering fractions determined from VLBA
mapping (see Section \ref{VLBA_sec}).  }
\end{table}
\end{center}

\subsection{Metallicities and spin temperatures}

Metal column densities were determined from the echelle spectra
using the apparent optical depth method on unsaturated transitions
(Table \ref{coldens_tab}).  The metal column densities were combined
with \nhi\ following standard procedures to determine an abundance relative
to solar, with solar abundances taken from Asplund et al. (2009):

\begin{equation}
[X/H] = \log N(X) - \log N(H) - \left[ \log(X)_{\odot} - \log (H)_{\odot} \right].
\end{equation}

We follow the standard assumption that each element's absorption is
dominated by a single ionization state such that N(Fe) = N(\FeII) etc.
This should be a good apprioximation for DLA gas shielded from the UV
background and whose ionisation state is determined by its stellar
population.  Final abundances are given in Table \ref{abund_tab}.

For \hii\ absorption studies of DLAs towards radio-loud QSOs, the
\hii\ optical depth, \hi\ column density \nhi\ and spin temperature
T$_s$ are related by the expression

\begin{equation}
\label{ts_eqn}
N(HI) = 1.823 \times 10^{18} \left[ \frac{T_s}{f} \right] \intop \tau {\rm d}V,
\end{equation}

where \nhi\ is in cm$^{-2}$, T$_s$ is in K, and d$V$ is in \kms. We
adopt the measurements of \nhi, $\tau$ and $f$ derived in Sections
\ref{GBT_sec}--\ref{echelle_sec} in the final estimates of our spin
temperatures.  In Table \ref{summary_tab}, we provide the final spin
temperatures (corrected for the source covering factor) for five of
the targets of our sample. Note that the table does not list a spin
temperature for the sub-DLA towards B0801+303; see Section
\ref{0801_sec} below.  In Figure \ref{21cm_echelle} we overlay the
\hii\ and metal line profiles.  In all cases, we show the same two
metal line transitions: \FeII\ $\lambda$2260 and \MgI\ $\lambda$ 2852.
Whilst \MgI\ is likely to be a better tracer of the cold neutral
medium (CNM) gas that is associated with \hii\ absorption than the
singly ionised species that are the most commonly detected
(e.g. \MgII, \FeII\ etc.), it can also have contributions from warm
ionised gas (e.g. Kanekar et al. 2010).  However, in the absence of,
for example, \CI\ detections, \MgI\ is our best tracer of the CNM
amongst the UV resonance lines.

Below, we briefly summarise the noteworthy features of each absorber.

\begin{figure*}
\centerline{\rotatebox{270}{\resizebox{14cm}{!}
{\includegraphics{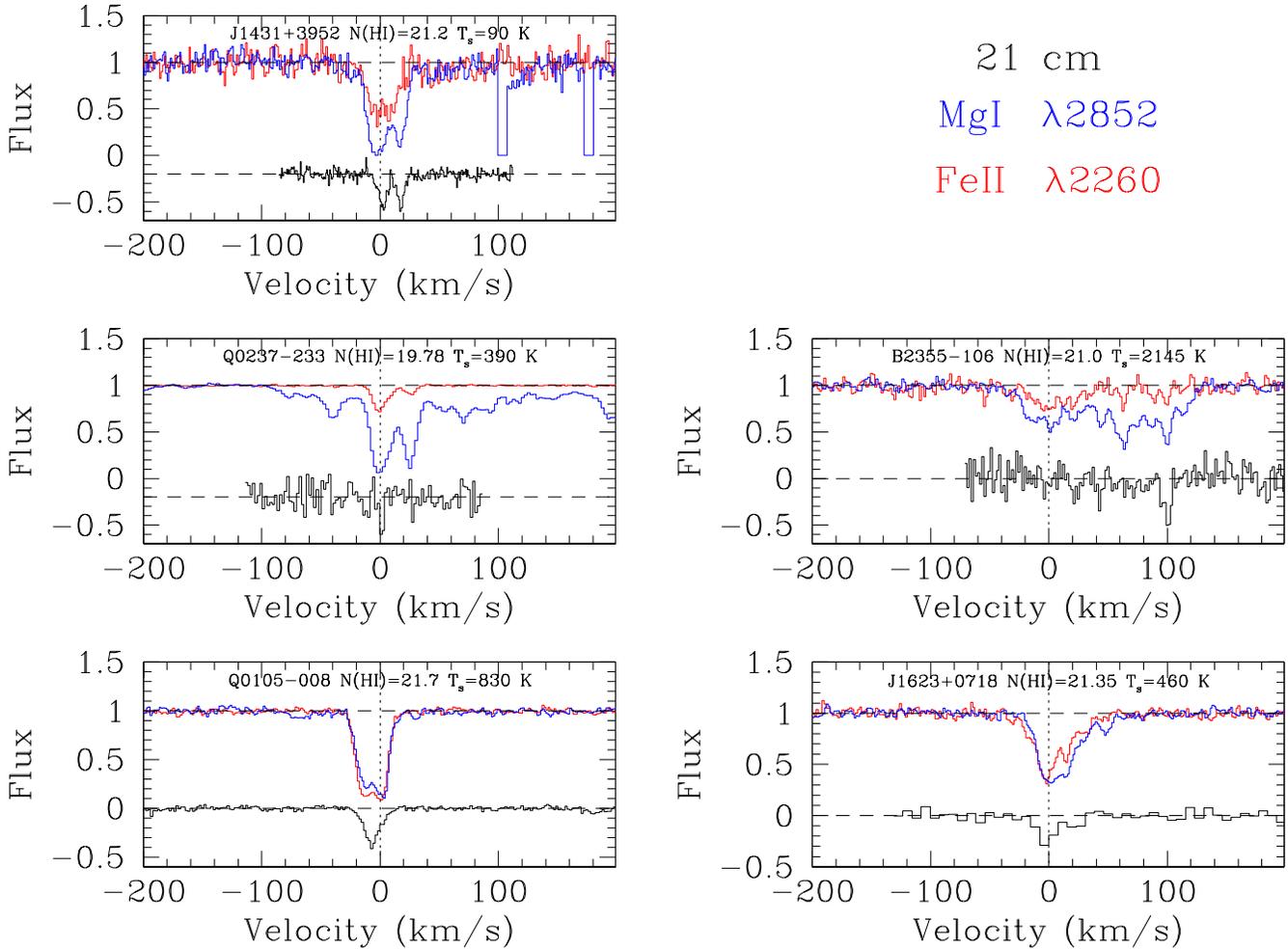}}}}
\caption{\label{21cm_echelle} Velocity profiles for the five absorbers
in our sample for which we have both \hii\ and echelle spectra. In
each panel, we show the same transitions on the velocity scales used
in Figures \ref{UM305_metals} -- \ref{2355_metals}. The \hii\
profiles are arbitrarily offset and scaled for display purposes.}
\end{figure*}

\subsection{Notes on individual absorbers}\label{indiv_sec}

The \hii\ absorbers presented in this paper are part of two samples 
of ``strong'' \MgII\ absorbers, selected from the SDSS and the CORALS 
survey (Ellison et al. 2001), towards radio-loud quasars that were 
surveyed for \hii\ absorption with the GMRT and the GBT (Kanekar et al.
2009b; Gupta et al. 2009). The selection criteria were 
(1)~$W_0^{\lambda 2796} \ge 0.5$ \AA\ and $W_0^{\lambda 2600} \ge 0.34$ \AA\ 
(where $W_0^{\lambda 2796}$ and $W_0^{\lambda 2600}$ are the rest equivalent 
widths in the \MgII$\lambda$2796 and \FeII$\lambda$2600 transitions, 
respectively) and (2)~that the absorber should lie towards a radio-loud 
background quasar with flux density $\gtrsim 300$~mJy at the redshifted 
\hii\ line frequency.

\subsubsection{B0105$-$008}

The relative velocity between the DLA and the background QSO is 
$\Delta v \sim$ 200 \kms.  Although this velocity offset is likely to be
uncertain by several hundreds of \kms\ (van den Berk 2001; Richards et
al. 2002; Shen et al. 2007), the absorber is almost certainly a ``proximate'' 
DLA (PDLA; e.g. the surveys by Ellison et al. 2002; Russell, Ellison \& 
Benn 2006; 
Prochaska, Hennawi \& Herbert-Fort 2008).  There is evidence that 
high \nhi\ PDLAs may have elevated abundances, by a factor $\sim 3$ 
(Ellison et al.  2010, 2011). Further, the spin temperatures in PDLAs are 
likely to be affected by the proximity of the gas to a bright radio 
source (e.g. Field 1958).  We discuss the treatment of PDLAs in
our sample explicitly in Section \ref{pdla_sec}.

The \hii\ absorption towards this absorber has a narrow profile 
that is well fit with a simple 2-gaussian model (Gupta et al. 2007; 
Kanekar et al. 2009b). As noted in Section \ref{VLBA_sec}, there 
is evidence of a core-jet structure in the 606~MHz VLBA image, which 
leads to a core covering factor of 0.32. The core is unresolved
along the declination axis, with angular extent $<< 18$~mas (i.e. 
spatial extent $<< 150$~pc).

The \hi\ column density in this absorber has a larger uncertainty than 
that of the other DLAs in our sample as the absorption overlaps with the
\lya\ emission line; this is a common problem for PDLAs.  Nonetheless,
the \nhi\ is clearly extremely high in this absorber, the largest in our 
sample with log~\nhi\ = $21.70 \pm 0.15$.

Both HIRES and UVES data are available for this sightline; the latter
were obtained from the UVES archive, from proposal 082.A-0569 (PI Srianand).
The UVES data have a higher S/N and wider wavelength coverage and are used 
in preference to the HIRES data for most transitions.  However, the HIRES data 
have more extensive coverage of the blue spectrum for transitions with rest 
wavelengths 1290~$< \lambda_0 <$1475 \AA.

The combination of good S/N and high \nhi\ facilitates the detection
of a wide range of metal line transitions, which all show a blended
two-component structure.  The \hii\ absorption is offset from the
strongest metal line component, and lies between the two components in
velocity space (Figure \ref{21cm_echelle}).  The column density of
\SiII\ is reported as a lower limit due to the mild saturation of the
$\lambda$1808 transition.  Overall, this absorber has a metallicity of
$Z \sim 1/25 Z_{\odot}$, which is relatively metal-poor for this
redshift (see Section \ref{results_sec}).  The low metallicity also
puts this absorber at odds with the majority of other high \nhi\
PDLAs, which tend to have relatively high abundances (Ellison et al.
2010, 2011).

\subsubsection{B0237$-$233}

The UVES spectrum of B0237$-$233, obtained as part of a UVES Large
Program, (166.A-0106, PI Bergeron), is truly exquisite.  The
combination of very long integration times (approximately 25 hours
split between two different wavelength settings) and a bright apparent
magnitude ($B=16.8$) leads to S/N ratios up to 150.  The high quality
of the UVES spectrum, which extends as blue as 3050 \AA, therefore
permits a precise measure of \nhi\ in this low $z$ absorber without
the need for HST.  We note the presence of a saturated \lya\ line in
the red wing of the DLA. Lack of coverage of unsaturated higher order
Lyman lines precludes an accurate \nhi\ measurement of this line.
However, this contamination is unlikely to affect the DLA fit, which
is driven by the extended damping wings.

Despite the low \nhi\ of this absorber, which is below the traditional
threshold of DLAs, the metal lines are sufficiently strong to enable
us to determine a wide suite of abundances.  The only species listed in
Table \ref{abund_tab} that we do not significantly detect is the
challenging \TiII.  There is actually a very weak absorption feature
at the expected wavelength of \TiII\ $\lambda$ 1910.6, but its significance
is only 2$\sigma$.  We therefore conservatively report the formal
3$\sigma$ limit for this transition.

The \hii\ absorption is well aligned with the strongest of the two
metal line components (Figure \ref{21cm_echelle})\footnote{The
extended absorption around the \MgI\ profile is due to contamination
from the atmospheric A band.}.  There is a single pixel feature in the
\hii\ absorption spectrum which is aligned with the weaker metal feature.
Although suggestive, this feature is not statistically significant in the
current GBT spectrum and is not included in the integrated \hii\ optical
depth.

The VLBA map of this quasar reveals a compact radio core, from which we 
derive a very high covering factor of 0.90. The spatial extent of the 
core emission is $\sim 97 \times 83$~pc$^2$, at the absorber redshift.

\subsubsection{B0801+303}\label{0801_sec}

Kanekar et al. (2009b) discuss the likely low covering fraction of
this absorber (see also Kunert et al. 2002). Our VLBA imaging shows a
weak, compact continuum source with a flux density that is only about
2\% of the total flux density measured with a lower resolution
interferometer. This clearly suggests that the radio emission from
this source is extended on scales larger than 0.22 arcsec; a value set
by the VLBA short spacing limit at 606 MHz. It is therefore likely
that the \hii\ absorption arises against the extended radio flux and
not against the radio core.  The likely non-coincidence of the optical
and radio sightlines implies that we cannot determine the spin
temperature of this absorber.  We therefore do not include this
sub-DLA in our spin temperature analysis (it is excluded from Table
\ref{summary_tab}) and have not sought to obtain high resolution
optical spectroscopy.

\subsubsection{J1431+3952}

This paper presents a new detection of \hii\ absorption in the
sightline towards J1431+3952. The integrated \hii\ optical depth is
quite high, with the profile consisting of 2 narrow absorption
components separated by $\sim$ 15 \kms\ (Figure
\ref{fig:J1431_GBT}). The two \hii\ absorption components are well
aligned with the velocity structure seen in \MgI\ (Figure
\ref{21cm_echelle}). The relatively low S/N around the \FeII\ $\lambda$
2260 line makes its velocity structure harder to discern.  However,
the two component structure is still present in the singly ionised species,
e.g. in \ZnII\ $\lambda$ 2026, again well aligned with the \MgI.

The VLBA image of J1431+3952 shows a core-jet structure, with a clear 
southern jet. The central radio emission was hence fit with a 2-component
gaussian model and a covering factor of 0.32 was obtained for the 
unresolved core. The core is extremely compact, with a deconvolved 
angular size of 0~mas along the declination axis, indicating that the 
angular extent along this axis is $<< 5.3$~mas (i.e. a spatial extent 
of $<< 35$~pc).

After correction for its sub-unity ($\sim 30$\%) covering factor, this
DLA has the lowest measured spin temperature to date, \ts$= 90 \pm
23$\,K. The only other DLAs with spin temperatures $< 150$\,K (after
correcting for the covering factor) are the systems at $z = 0.395$
towards Q1229$-$021 (Brown \& Spencer 1979) and $z = 2.289$ towards
TXS0311+430 (York et al. 2007) which have \ts=95 and 120 K
respectively.  In the measurements of Galactic spin temperatures by
Kanekar et al. (2011), two Milky Way sightlines have comparably low
values (\ts=89 and 119\,K), both at very low Galactic latitudes.
Given the suggestion that DLAs with low spin temperatures are often
(at least at low $z$) associated with spiral galaxies (Chengalur \&
Kanekar 2000; Kanekar \& Chengalur 2001), the high implied CNM
fraction towards J1431+3952 may be associated with the intersection of
a galactic disk.

The SDSS image of J1431+3952  shows a faint galaxy at $\sim$ 5 arcsec
separation from the QSO.  At a redshift $z=0.602$ this would correspond
to a physical projection of $\sim$ 38 \hkpc, which is fairly large
given the implied intersection of a cold disk gas.  Moreover, a fit
of the SDSS photometry yields a photometric redshift $z=0.08\pm0.02$
(Michael Palmer, 2012, private communication).
A high spatial resolution image of the J1431+3952 field would be of great
interest to attempt to identify the absorbing galaxy.

\subsubsection{J1623+0718}

The redshift of the \hii\ absorption ($z=1.33567$) is in excellent
agreement ($\sim 4$ \kms) with the centroid of the metal lines ($z=1.33570$).
For the majority of metal line transitions, we use the UVES spectrum,
for its superior S/N.  However, the \ZnII\ $\lambda$ 2026 line falls
between the blue and red arm coverage, so the HIRES data is used for
this transition.  The overall metallicity of this absorber is $Z \sim
1/10 Z_{\odot}$.

In order to compare the structure in the metal lines with the \hii\
absorpion (originally presented by Gupta et al. 2009), we have
re-reduced the archival GMRT data (project 14RSA01, PI Srianand) for
this target using standard procedures in AIPS (see, for example,
Kanekar et al. 2009b for a full description of the analysis procedure of
610-MHz GMRT data). Our re-reduction of the GMRT data is presented
in Figure \ref{21cm_echelle}; the strongest \hii\ absorption is
well-aligned with the strongest metal component.

The VLBA image of J1623+0718 shows a strong core, with a possible weak
jet, only tentatively detected in the present map. Only one-third of
the total radio emission stems from the core, with a covering factor
of $\sim 0.34$; most of the emission must arise from extended
structure that is not detected in the VLBA image. The spatial extent
of the core emission is quite small, $\sim 36 \times 12$~pc$^2$, at
the absorber redshift.

\subsubsection{B2355$-$106}

The DLA towards B2355$-$106 has \ts=2145 K, making it one of the
highest spin temperatures to date in a DLA with a detection of \hii\
absorption.  The \hii\ absorption reported by Kanekar et al. (2009b)
in this DLA is located in a single narrow component at $z=1.17303$.
In contrast, the metals show absorption over $\sim$ 150 \kms\ (Figure
\ref{2355_metals}), with the strongest \FeII\ absorption at $1.17230$,
offset from the \hii\ absorption by $\sim 100$~\kms.  However, \MgI\
is strongest in the redder components (at $v > 0$ \kms\ in Figure
\ref{2355_metals}), which matches well the \hii\ absorption (Figure
\ref{21cm_echelle}).  Although not statistically significant, there
may be some extended \hii\ absorption bluewards of the main detected
feature, spanning the broad velocity range over which the metal
absorption is observed (Figure \ref{21cm_echelle}).  However, the
extended absorption is not detected in the independent spectrum of Gupta
et al. (2009), so we do not include it in the integrated optical
depth and spin temperature determination.

There is possible \CrII $\lambda$ 2056 absorption, but the
significance of the feature is less than 3 $\sigma$, so a limit is
quoted for N(\CrII) in Table \ref{abund_tab}. The \TiII\ $\lambda$1910 line is
formally detected at 4.9 $\sigma$, but visually this is not very
convincing.  The significance of detections in this absorber are
complicated by the broad velocity structure, so we conservatively
quote the \TiII\ as a limit.  The same broad structure, combined
with a relatively low S/N, makes weak features particularly uncertain,
which may lead to over-estimation of column densities (e.g. \ZnII).

The VLBA image of this source shows an extremely compact radio core, 
with a spatial extent of $\sim 90 \times 18$~pc$^2$ at the absorber 
redshift, and a covering factor of unity.

\section{Results and discussion}\label{results_sec}

\subsection{Filling in the redshift desert}

\begin{figure}
\centerline{\rotatebox{270}{\resizebox{6.5cm}{!}
{\includegraphics{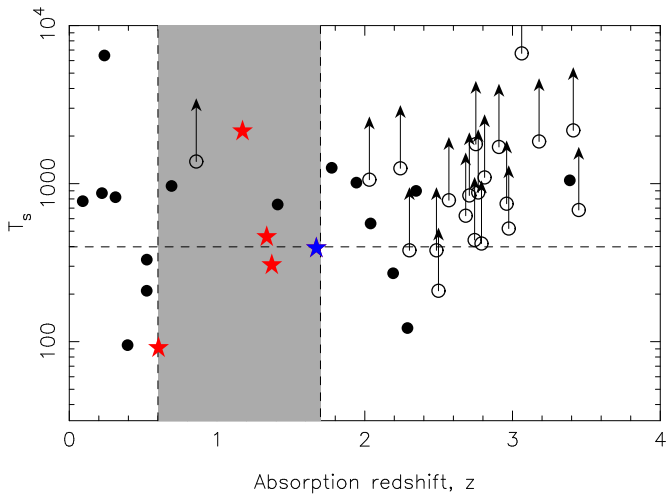}}}}
\caption{\label{ts_z} Spin temperature as a function of redshift for
all sub-DLAs and DLAs (including proximate absorbers) with measured
covering factors.  Filled and open circles represent \hii\ detections
and limits from the literature respectively.  Our intermediate
redshift sample is shown by filled stars in the redshift desert ($0.6
< z< 1.7$, shaded region). Red stars indicate DLAs and blue stars show
sub-DLAs.  The horizontal dashed line shows a spin temperature of 400
K. }
\end{figure}

The measurements presented in this paper are combined with a wider
literature sample of \ts\ measurements in order to provide the first
attempt at bridging the high and low redshift observations and to
improve the overall statistics of spin temperatures and metallicities.
The spin temperatures of DLAs and sub-DLAs are shown as a function 
of redshift in Figure \ref{ts_z}. All spin temperatures in this figure 
include the estimates of the DLA covering factor; indeed, in the 
following sections, we will deal only with $f$-corrected spin 
temperatures unless otherwise stated\footnote{We use only covering
factors that have been determined at frequencies within a factor of
$\sim 2$ of the \hii\ absorption}.  In a companion paper (Kanekar et al. in
preparation), we will present the full DLA sample, including both new 
GBT/GMRT \hii\ absorption studies and sources from the literature, 
as well as the results of an extensive VLBA campaign to determine 
covering factors for many of the absorbers presented in Figure~\ref{ts_z}. 
The redshift desert between $0.6 < z_{\rm abs} < 1.7$ is obvious in 
Figure~\ref{ts_z} -- prior to the work presented
here, only 2 \hii\ detections existed in this range.  We have added a
further 5 \ts\ measurements (filled stars) which sample the regime
over which the spin temperature has been claimed to evolve.  It is also
interesting to note that our sample encompasses absorbers with a wide
range of spin temperatures, including the lowest, and one of the
highest values amongst \hii\ detections.  The full analysis of the
spin temperature evolution will be presented by Kanekar et al. (in
preparation).  For the remainder of this paper, we focus on the
metallicity properties of \hii\ absorbers.

\subsection{Proximate DLAs and sub-DLAs}\label{pdla_sec}

Four of the DLAs shown in Figure \ref{ts_z} are proximate DLAs within
3000 \kms\ of the QSO (B0105$-$008, Q0405$-$331, Q1013+615 and
Q0528$-$250), including one of the absorbers in our intermediate
redshift sample (B0105$-$008).  There are various reasons why PDLAs
might be excluded from a general analysis of abundances in \hii\
aborbers.  First, radiation from the QSO may alter the hyperfine 
level populations in the \hi\ atoms (e.g. Field 1958; Wolfe \& Burbidge 
1975). The result of such irradiation would be that the observed spin 
temperature would depend both on the radiation field and on the 
distribution of gas in different temperature phases, significantly 
complicating the interpretation. Three of the four PDLAs in our sample 
have \ts$ > 750$\,K; they are all at $z_{\rm abs} > 2.5$ where such spin
temperatures are typical (Figure \ref{ts_z}).  The final PDLA is the
absorber towards Q0105$-$008 from our redshift desert sample ($z_{\rm
abs} = 1.37$), with \ts\ = 305\,K (including the correction for a covering 
factor of 0.32).

Independent of the \hii\ absorption, it has been suggested that
PDLAs with log \nhi\ $> 21$ are biased towards higher metallicities
(Ellison et al. 2010, 2011).  Combined with a possible anti-correlation 
between \ts\ and metallicity, the high \nhi\ PDLAs may have preferentially 
low values of \ts.  Two of the PDLAs in the sample, B0105-008 and 
Q0528-250, have log \nhi~$> 21$. The abundances of B0105-008 are 
presented here, and we find a very typical metallicity for this redshift, 
[Zn/H]=$-1.40$.  Q0528-250 was included in the PDLA study of Ellison 
et al. (2010) and does show a relatively high metallicity for its 
\nhi\ (see Figure 15 in Ellison et al. 2010).

In view of these potential biases, we limit the following analyses
to absorbers that are at least 3000~\kms\ from the published quasar
systemic redshift.    

Another class of absorbers that is often distinguished from the general DLA 
population consists of the so-called sub-DLAs, systems with 
$19.0 < $~log~\nhi~$< 20.3$.  Although their \nhi\ falls below the classical 
DLA criterion, they show clear damping wings and in many cases have
reliable abundances (e.g. Dessauges-Zavadsky et al. 2003, although see
Richter et al. 2005; Quast et al. 2008; Milutinovic et al. 2010 for
counter-examples).  Although Peroux et al.  (2003) have suggested that
sub-DLAs make an important contribution to cosmic \hi, particularly at 
high redshifts, O'Meara et al. (2007) calculated that systems with 
log \nhi~$< 20.3$ represent only $\sim$ 10\% of cosmic neutral hydrogen. 
Prochaska, Herbert-Fort \& Wolfe (2005) have also argued that the 
sub-DLAs are not tracers of the main gas reservoirs for cosmic star 
formation.   Indeed, there is evidence from Galactic sightlines that 
below the DLA \nhi\ threshold there is only a small contribution from 
the CNM (Kaneker et al. 2011).  Finally, it is observed that low-redshift 
sub-DLAs (which have been mostly identified through their large 
\MgII\ equivalent widths) have systematically higher metallicities than 
DLAs at the same redshift (e.g. Peroux et al. 2008; Meiring et al. 2009;
Dessauges-Zavadsky et al. 2009). 

Despite these literature debates, we do include sub-DLAs in the
current work, as it is of interest to see whether the Galactic trends
with \nhi\ are also evident in cosmological DLAs and sub-DLAs. In
particular, if the CNM fractions (and hence, spin temperatures) depend
on metallicity, we might expect trends between these parameters to
extend into the sub-DLA regime, regardless of the nature of the
intervening galaxy.  However, we do distinguish sub-DLAs from DLAs in
figures and quoted statistics.

\subsection{Metallicities in \hii\ absorbers}

\begin{figure*}
\centerline{\rotatebox{0}{\resizebox{16cm}{!}
{\includegraphics{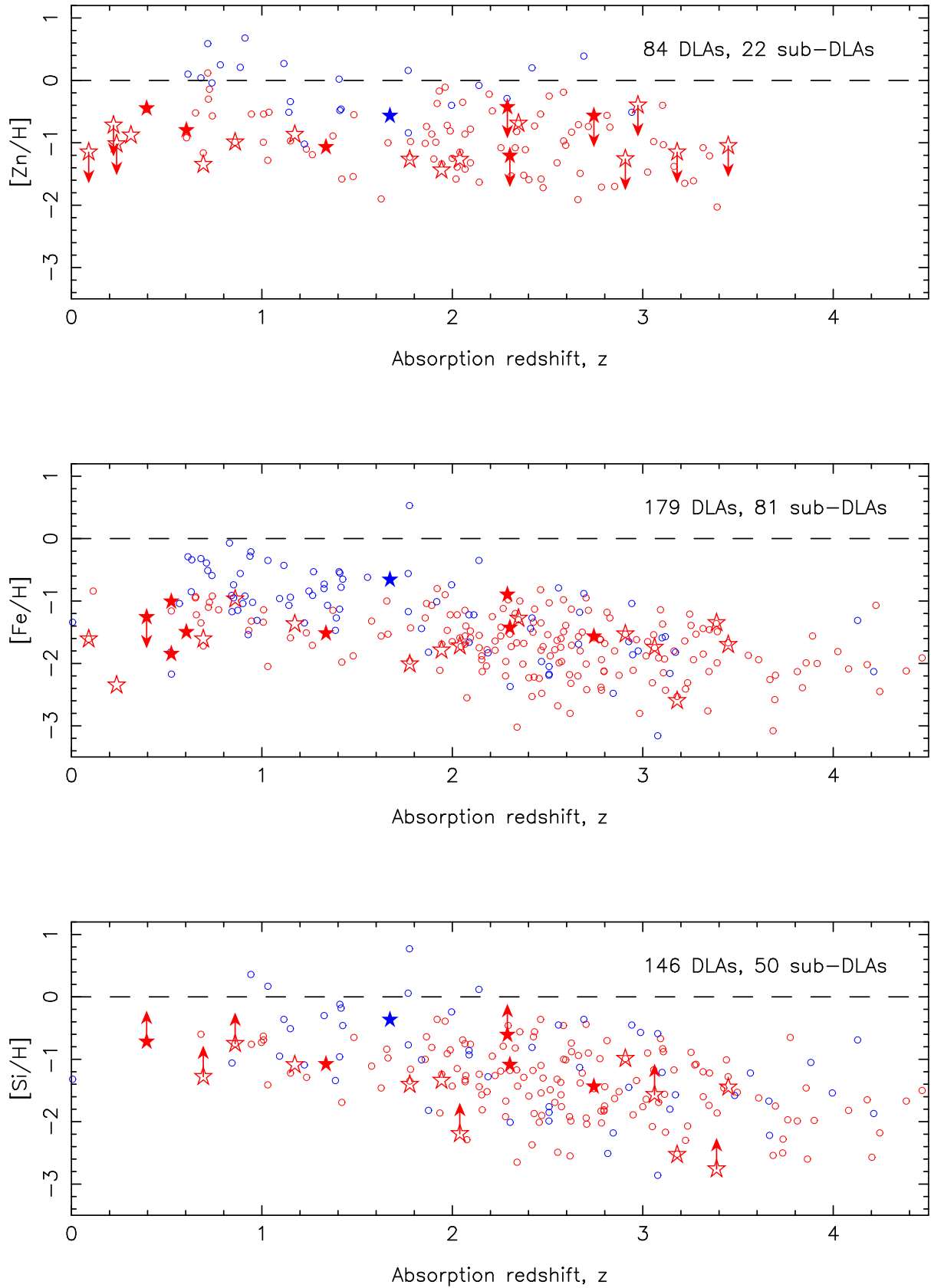}}}}
\caption{\label{abund_z}Abundances taken from the literature for DLAs
and sub-DLAs (red and blue open circles respectively).  For clarity,
limits are not shown for the literature sample of DLAs and sub-DLAs
for which there are no \hii\ observations.  \hii\ absorbers with
measured covering factors are shown with stars, again coloured coded
accorded to their classification as DLAs (red) or sub-DLAs (blue).
Open stars show \hii\ detections and limits with \ts~$> 500$\,K and
filled stars are for \hii\ absorbers with \ts~$< 500$\,K.  }
\end{figure*}

In Figure \ref{abund_z} we show the abundances of three commonly
measured elements (Zn, Fe and Si) in DLAs and sub-DLAs with \hii\
absorption studies, as a function of redshift.  We distinguish low and
high (including lower limits) spin temperatures by plotting
filled stars for absorbers with \ts\ $<$ 500 K and open stars
for absorbers with higher values of \ts.  For comparison, we show a
literature compilation (Berg et al, in preparation; Rafelski et al.
2012) from DLAs and sub-DLAs (open circles) towards 
all quasars (i.e. not merely radio quasars). DLAs and sub-DLAs 
for both samples are shown in red and blue, respectively.

The now well-documented (e.g. Peroux et al. 2008; Meiring et al. 2009)
enhancement in sub-DLA abundances is clearly present at low redshifts
($z_{\rm abs} < 1.7$).  Whilst it has been suggested that their high
metallicities indicate that sub-DLAs sample the most massive galaxies,
Dessauges-Zavadsky et al. (2009) have argued that this may be a
selection effect, since the majority of sub-DLAs in this redshift
range were selected via the \MgII-based survey of Rao et al. (2006).

Whilst \MgII\ selection seems to sample DLAs in an unbiased way,
Dessauges-Zavadsky et al. (2009) have argued that at lower \nhi\ there
may be a bias towards larger velocity widths and/or metallicities (see
also Rao et al. 2006 for a discussion on selection bias, and Bouche 2008
for further evidence of a bias towards high metallicity sub-DLAs selected
by \MgII).  In
either case, the sample selected using \hii\ absorption (many of which, 
at $z_{\rm abs} < 1.7$, were also originally selected based on \MgII\ rest 
equivalent width) show evidence of the same metallicity trends as the 
absorbers towards the full quasar sample. The \hii\ absorber metallicities 
cover the same ranges as those of the full sample, and the \hii\ sub-DLAs show
evidence for slightly higher metallicities than DLAs at the same
redshift.

Although the elevated metallicities in \hii\ sub-DLAs are consistent
with optical samples, known trends in \hii\ absorbers may make the
high metallicities in sub-DLAs with \hii\ absorption seem puzzling.
Kanekar et al. (2011) have shown that Galactic sightlines with \nhi\
in the sub-DLA regime tend to have high spin temperatures, and
Kanekar et al. (2009c) find that high spin temperatures are more usually
found in low metallicity systems.  So, we might expect that \hii\ absorbers
in the sub-DLA regime should have relatively low (rather than the observed 
high) abundances.

\begin{figure}
\centerline{\rotatebox{270}{\resizebox{6cm}{!}
{\includegraphics{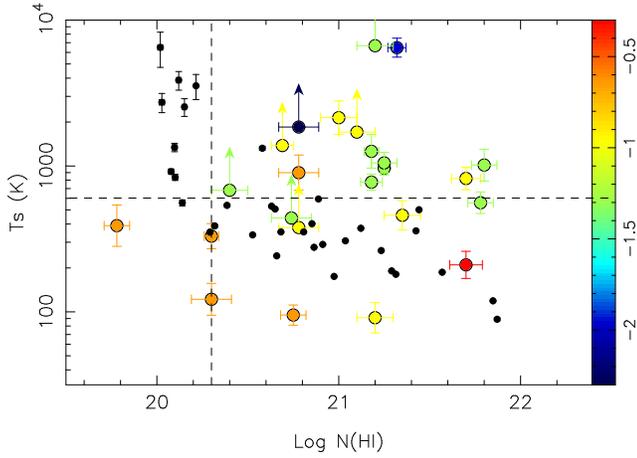}}}}
\caption{\label{HI_ts} Covering factor-corrected spin temperatures
versus \nhi\ for DLAs, colour-coded by metallicity
([Zn,S,Si,Fe+0.4/H]).  PDLAs are excluded from this Figure.  For
comparison, Galactic sightlines taken from Kanekar et al. (2011) are
shown with smaller black points. The horizontal dashed line shows a
spin temperature of \ts~$= 600$\,K.  The vertical dashed line indicates the
division between sub-DLAs and DLAs (at log \nhi~$ = 20.3$).}
\end{figure}

In Figure \ref{HI_ts} we put these arguments on a quantitative
footing by showing the spin temperature and \nhi\ Galactic
data points used by Kanekar et al. (2011), as well as the values for
\hii\ detections in DLAs. The latter have been colour-coded by metallicity.
Metallicities are taken to be, in order of preference, [Zn/H], [S/H],
[Si/H], [Fe/H]+0.4 (where the offset accounts for typical dust
depletions; Prochaska \& Wolfe 2002).  Since the correction to [Fe/H]
is rather crude we give priority to, for example, a [Zn/H] limit over
a [Fe/H] detection.  The exception to these priorities is AO\,0235+164
for which we use an X-ray metallicity, as there is evidence that this
is a very dusty absorber with depletion considerably larger than the
typical value of 0.4 that we otherwise adopt (e.g. Junkarrinen et al 2004;
Kulkarni et al. 2007).  The lower left part of the
diagram with \nhi $< 2\times10^{20}$ \cm\ and \ts$<600$ K is poorly
populated by the Galactic and sub-DLA sightlines alike.  This supports
the suggestion by Kanekar et al. (2011) that below this column density
threshold, the neutral medium is dominated by warm gas.  At higher
column densities, the DLAs show much more scatter in \ts\ than the
Galactic data, but with a clear dependence on metallicity.  The DLAs
with metallicities below around 1/10 of the solar value (yellow, green
and blue points) have high spin temperatures for their \nhi, relative
to the Milky Way points.  The more metal-rich DLAs (orange and red
points) have either comparable, or low spin temperatures for their
\nhi.  Such a metallicity dependence may be expected from cooling
arguments, whereby metals aid the cooling of the neutral medium 
(Wolfire et al. 1995; Kanekar \& Chengalur 2001), in a similar way 
that metals have been suggested to aid the transformation from the 
neutral to molecular regimes (Schaye 2001; Krumholz et al. 2009).

\begin{figure}
\centerline{\rotatebox{270}{\resizebox{6cm}{!}
{\includegraphics{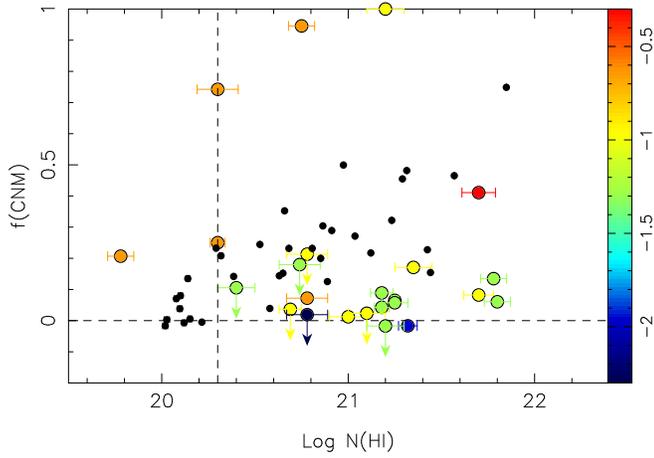}}}}
\caption{\label{HI_fCNM} The fraction of gas in the CNM versus \nhi\
for DLAs, colour-coded by metallicity ([Zn,S,Si,Fe+0.4/H]).  PDLAs are
excluded from this Figure.  For comparison, Galactic sightlines taken
from Kanekar et al. (2011) are shown with smaller black points.  The
vertical dashed line indicates the division between sub-DLAs and DLAs
(at log \nhi~$ = 20.3$).}
\end{figure}

In Figure \ref{HI_fCNM}, we make a crude conversion of the spin
temperature into a CNM fraction (see also Kanekar \& Chengalur 2003).
For this calculation, in which we adopt a two phase medium, we must
assign spin temperatures to the cold and warm neutral media (WNM).
For the CNM, the spin temperature is the same as the kinetic
temperature (e.g.  Liszt 2001), and we adopt a value of
90K\footnote{Although CNM temperatures of 100--200 K are often assumed
(e.g. Kanekar \& Chegalur 2003), the lowest \ts\ in our sample is 90K.
So as not to exceed CNM fractions of unity, we adopt 90\,K as the CNM
temperature.}.  However, in the WNM, the low densities are
insufficient to thermalise the \hii\ transition (e.g. Liszt 2001).
Therefore, although the kinetic temperature of the WNM is expected to
be 5000--8000\,K, the spin temperature, which will depend on the
physical conditions within the cloud, is expected to be considerably
lower, 1000-4000\,K (Liszt 2001).  Although the CNM fractions that are
calculated will depend on the temperatures used, the qualitative
distribution of points remains the same.  We adopt a \ts\ for the WNM
of 3000\,K, although we note that this will lead to an (unphysical)
negative CNM fraction if the measured \ts $>$ 3000 K.  Figure
\ref{HI_fCNM} shows that, as suggested by Kanekar et al. (2011), the
low \nhi\ Galactic sightlines are dominated by the WNM.  As the \nhi\
increases, a higher CNM fraction is present. The sub-DLA in our sample
also has a relatively low CNM fraction, consistent with the upper end
of the distribution seen in Galactic (sub-DLA) sightlines. In the DLA
regime, the low metallicity absorbers have relatively low CNM
fractions, compared with the Milky Way.  The DLAs with high CNM
fractions tend to be relatively metal-rich.

\subsection{The spin temperature-metallicity anti-correlation}

\begin{figure}
 \centerline{\rotatebox{0}{\resizebox{8cm}{!}
{\includegraphics{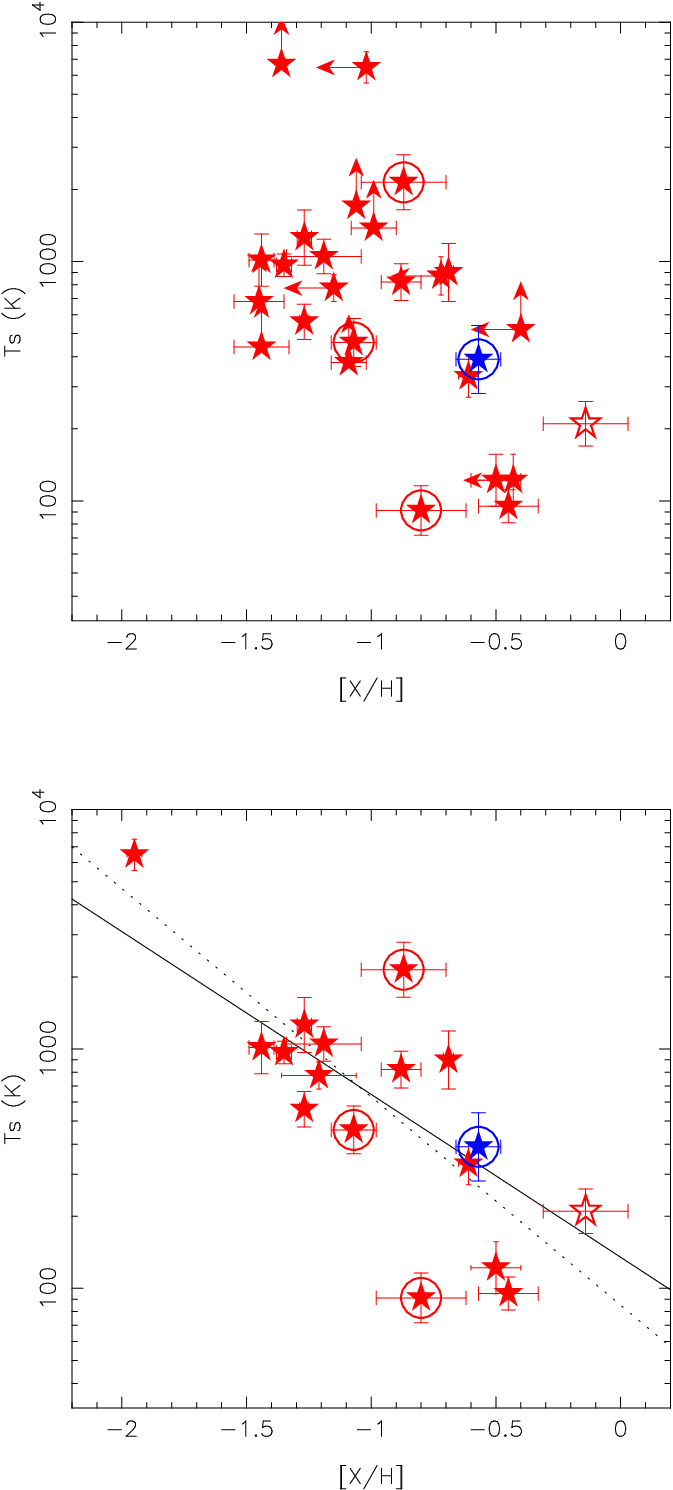}}}}
\caption{\label{ts_met} Spin temperature as a function of metallicity ([Zn,S,Si,Fe+0.4/H])  for all DLAs (red) and sub-DLAs (blue)
with measured covering factors.  Absorbers from our new sample (this
paper) are circled.  The open star represents AO\,0235+164 for which we
adopt the X-ray metallicity from Junkkarinen et al. (2004).  The upper
panel includes metallicity and/or spin temperatures limits.  The lower
panel includes only absorbers for which both \hii\ and metal
absorption is detected.  The dotted line shows the best fit for the
detections taken from Kanekar et al. (2009), whereas the solid line
shows the best fit for the full data sample presented here}
\end{figure}

In the previous sub-section we showed that sub-DLAs tend to have high \ts\ (and
therefore high fractions of warm gas), regardless of their often elevated
metallicities.  DLAs \textit{may} exhibit significant fractions of
cold gas, but this is likely to be metallicity dependent.  

We now turn to the explicit dependence of spin temperature on
metallicity.  Kanekar \& Chengalur (2001) predicted that an
anti-correlation between \ts\ and metallicity would naturally arise if
the paucity of cold neutral gas in DLAs was due to fewer cooling
routes due to a lack of metals in the absorbers.  The largest sample
to have been previously used to address this issue was a sample of 26
absorbers, of which 10 systems had both \ts\ and metallicity
measurements and 20 systems had estimates of the low-frequency
covering factors (Kanekar et al. 2009c). The number of detections in
our sample is significantly increased to 17, even with the additional
requirement that all absorbers must have low frequency covering factor
measurements to be included in our sample.  The new data presented
here include DLAs with very high and low spin temperatures, which are
particularly useful for anchoring any relationship with metallicity
(Figure \ref{ts_met}).  We note that the lowest metallicity absorber
in the lower panel of Figure \ref{ts_met} (which has the highest spin
temperature amongst the systems with both \hii\ and metal line
detections) is Q0952+179.  Due to our prioritization of metallicities,
this DLA appears in the upper panel as an upper limit from [Zn/H], but its
measured value of [Fe/H] is used in the lower panel.

\medskip

For the full sample of 26 detections and limits, a non-parametric
generalised Kendall-tau rank correlation test (the BHK statistic in
the {\sc ASURV} package; Brown et al. 1974; Isobe et al. 1986) detects
an anti-correlation between metallicity and spin temperature at $3.4
\sigma$ significance (Figure \ref{ts_met}, upper panel), treating the
metallicity as the independent variable.  If the DLA towards
AO\,0235+164 is dropped from the sample (in contrast with the rest of
the sample, its metallicity has been obtained from a X-ray
measurement), the significance of the anti-correlation is $ 3.1
\sigma$. Reducing the sample to only the 17 systems with measurements
of both \ts\ and metallicity (i.e. excluding limits; see
Figure~\ref{ts_met}, lower panel) yields an anti-correlation with $3.0
\sigma$ significance. Again, dropping AO\,0235+164 from this detection
sample gives an anti-correlation with a $2.8 \sigma$ significance (16
systems).

There is only one sub-DLA included in Figure \ref{ts_met}, the $z_{\rm abs} 
= 1.6724$ absorber towards Q0237$-$233, whose spin temperature we present 
for the first time in this work.  Despite the low \nhi, the spin temperature 
is relatively low in this absorber: \ts$ = 390 \pm 127$\,K. The moderately 
high metallicity ($Z \sim 1/4 Z_{\odot}$) places this sub-DLA in good 
agreement with the DLA data on Figure~\ref{ts_met}.  Dropping the one 
sub-DLA from the sample yields a \ts-metallicity anti-correlation that 
is significant at $3.2\sigma$ and $2.7 \sigma$ for limits+detections and 
detections only, respectively.

We also used a linear regression analysis to obtain the best-fit
relation between spin temperature (log[\ts]) and metallicity [Z/H],
applying this to the 17 DLAs with measurements of (rather than limits
on) {\it both} \ts\ and [Z/H] (see Figure~\ref{ts_met}, lower
panel). We used the BCES(Y$|$X) estimator (Akritas \& Bershady 1996) for
this purpose, again treating [Z/H] as the independent variable X.
Note that this method takes into account measurement errors on both
variables, as well as correlations between these errors. This is
relevant here as both \ts\ and [Z/H] are derived from the \hi\ column
density, and the errors on the two quantities are hence
correlated. The regression analysis yielded log[\ts]~$= (-0.87 \pm
0.14) \times {\rm [Z/H]} + (1.93 \pm 0.19)$, consistent with the
results of Kanekar et al. (2009c). The fit is shown as a solid line
in the lower panel of Figure~\ref{ts_met}.  If, instead of the (Y$|$X)
technique, we adopt the BCES bisector method for independent variables,
the fit is consistent to within 1$\sigma$:  log[\ts]~$= (-1.11 \pm
0.16) \times {\rm [Z/H]} + (1.70 \pm 0.19)$.

\begin{figure}
\centerline{\rotatebox{0}{\resizebox{8cm}{!}
{\includegraphics{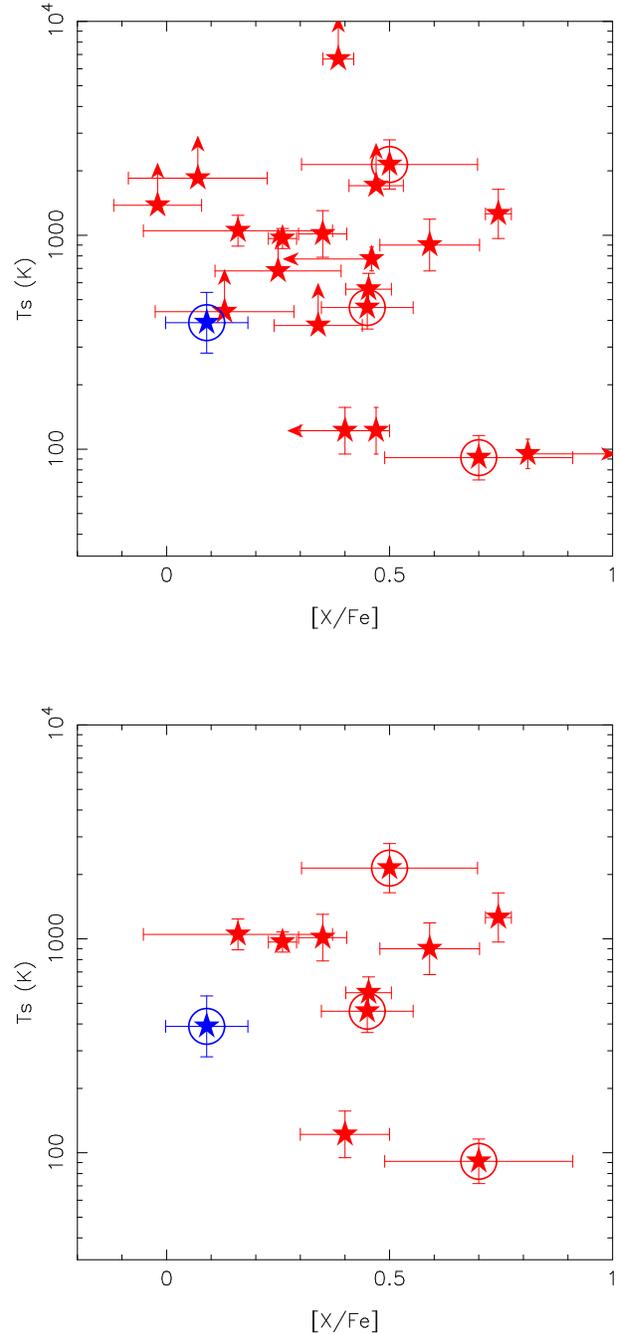}}}}
\caption{\label{dust_ts} Spin temperature as a function of dust
depletion as measured by [X/Fe]=[Zn,S,Si/Fe] for absorbers with
measured covering factors.  The upper panel shows detections and
limits in \hii\ and metals.  The lower panel shows only detections in
\hii\ and metals.  Sub-DLAs are shown in blue, DLAs are shown in red
and absorbers from our new sample (this paper) are circled. PDLAs are
excluded from this Figure.  }
\end{figure}

Kanekar et al. (2009c) did not find a significant anti-correlation
between dust depletion and \ts.  In Figure \ref{dust_ts} we confirm
this by plotting \ts\ versus dust depletion [X/Fe] (where X=Zn, S, Si,
or the X-ray value for AO\,0235+164) for our expanded sample. Applying
the generalised Kendall tau test to our sample of 22 systems with
either measurements of or limits on \ts\ and dust depletion (Figure
\ref{dust_ts}, upper panel) we find a $1.9 \sigma$ anti-correlation
between \ts\ and [X/Fe], considerably weaker than the anti-correlation
between \ts\ and metallicity\footnote{Adding a further two absorbers
to the depletion sample which do not have Fe abundances, but have
detections or limits in \CrII\ increases the significance negligibly
to 2.0 $\sigma$.}.  Note that only a handful of DLAs (11 systems) have
measurements of both dust depletion and \ts; this sub-sample shows no
evidence of an anti-correlation between [X/Fe] and \ts, with a
significance of $\sim 0.4\sigma$. The lack of a significant dependence
of \ts\ on depletion demonstrates that the \ts-metallicity
anti-correlation is not driven by an underlying dependence on dust. We
also note that the DLAs with \hii\ absorption show a range of
depletions that is well-matched to the range in the full DLA sample,
where the mean value of [Zn/Fe]$\sim$+0.4 (e.g. Prochaska \& Wolfe
2002).

\subsection{\hii\ absorption as a tool for selecting DLAs.}

\begin{figure}
\centerline{\rotatebox{270}{\resizebox{6cm}{!}
{\includegraphics{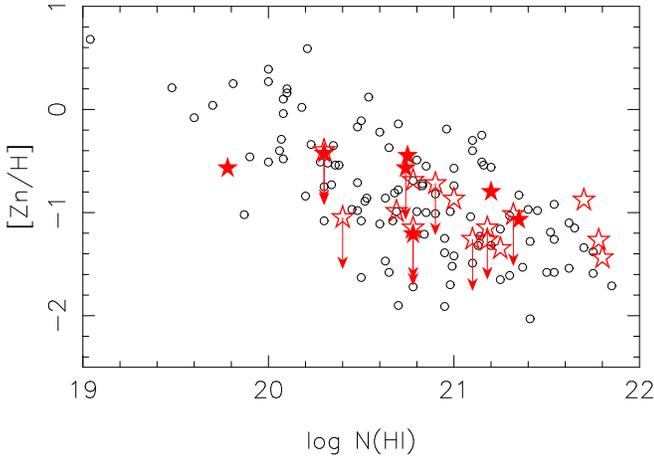}}}}
\caption{\label{HI_Zn} Metallicity, as traced by [Zn/H] as a function
of \nhi\ for a literature sample of optically selected DLAs (black open
circles) and \hii\ absorbers (red stars).  Open stars show \hii\
detections and limits with \ts $>$ 500 K and filled stars are for
\hii\ absorbers with \ts $<$ 500 K.  }
\end{figure}

The absorbers in the sample selected here were identified first as
\MgII\ systems, then as \hii\ absorbers, with the \nhi\ determination
as the final step.  Excluding B0801+303, which was \textit{post facto}
found to be an extended radio source, 4/5 of the \MgII-selected \hii\
absorbers in our sample are found to be DLAs.  This compares with a
35\% DLA identification rate for absorbers selected purely on \MgII\
equivalent width (Rao et al. 2006) and $\sim$ 60\% when velocity
widths are also taken into account (Ellison 2006; Ellison, Murphy \&
Dessauges-Zavadsky 2009).  The one sub-DLA in our main sample is
towards B0237$-$233, whose \hii\ optical depth is the lowest in our
sample by a factor of $\sim$ 3.

The connection between \hii\ absorbers and DLAs is not particularly
surprising.  Several works (e.g. Kanekar \& Chengalur 2003; Curran et
al. 2010) have noted a correlation between the integrated \hii\
optical depth and \nhi.  Further, Kanekar et al. (2011) show that most
Galactic sightlines with \nhi$<2\times10^{20}$ \cm\ have relatively
high spin temperatures, and are hence dominated by the WNM.  Only
above this approximate threshold can the neutral gas apparently cool
sufficiently to form a significant fraction of the cool phase, then
becoming readily detectable in \hii\ absorption.

\hii\ absorbers may therefore represent an interesting alternative
identification of DLAs at low redshifts.  The obvious concern is how
representative such a selection would be.  Given the results discussed
in this paper, it may be expected that \hii\ absorption is most readily
detected at large \nhi\ and high metallicity.  Indeed, all four DLAs of 
the present sample have high \nhi, in excess of 10$^{21}$ \cm.  
This is unsurprising, since for a given \ts\ the \hii\ optical depth 
will be highest for the highest \nhi\ absorbers.  Since DLAs with 
log~\nhi~$> 21$ are relatively rare, \hii\ selection offers a good 
opportunity to identify the highest column density absorbers for 
targeted studies.  For example, many of the `metal-strong' DLAs 
(Herbert-Fort et al. 2006; Kaplan et al. 2010), which are useful
for the identification of rare metal species, have such high \nhi.

Despite the bias towards high \nhi\ in an \hii\ selected sample of
DLAs, the metallicities and dust depletion in \hii\ absorbers span
the observed range in the full sample of DLAs and sub-DLAs (Figures
\ref{abund_z} and \ref{dust_ts}).  In Figure \ref{HI_Zn}, we also show
the metallicities (as measured by [Zn/H]) as a function of \nhi.
It has been argued that the anti-correlation between [Zn/H] and
\nhi\ is inconsistent with a dust bias, given the typical reddening
in DLA samples (Elllison, Hall \& Lira 2005).  The paucity of
high \nhi\ absorbers with high metallicities can instead be understood
as a metallicity-dependent transition from the atomic phase to a phase 
dominated by molecular gas (Schaye 2001; Krumholz et al. 2009).  The 
\hii\ absorbers have metallicities consistent with those of the full 
sample of DLAs at a given \nhi.  Combined with Figures \ref{abund_z} and 
\ref{dust_ts}, this indicates that an \hii\ selected sample is not strongly
biased for studies of elemental abundances.

\section{Conclusions}

The study of \hii\ absorbers at cosmological distances is now coming
of age in terms of sample sizes, redshift coverage and availability of
covering factors.  This paper focuses on our efforts to fill in the
spin temperature redshift desert ($0.6 < z < 1.7$) and investigate the
role of metallicity in modulating spin temperature trends.  In a
companion paper (Kanekar et al. in prep.)  much of the supporting high
redshift data used in our analysis is presented, as well as a
analysis of the redshift evolution of the spin temperature. The main 
results of the current paper are as follows.

\begin{enumerate}

\item For a sample of six \MgII-selected \hii\ absorbers in the
redshift range $0.6<z_{\rm abs}<1.7$ we have determined \nhi\ (mostly
from HST spectra), metal column densities from optical
echelle spectra and covering factors ($f$) from VLBA images.

\item One of the absorbers ($z_{\rm abs}=0.60190$ towards J1431+3952)
is a new \hii\ detection made with the GBT.  Its spin temperature is 
the lowest yet reported: \ts=90$\pm$23 K.

\item The metallicities and dust depletions of \hii\ absorbers span
the range that is typical for the full DLA sample in the same
redshift range (or at fixed \nhi).  As is the case for the full DLA, 
\hii\ sub-DLAs have slightly higher metallicities than \hii\ DLAs, 
at a fixed redshift.

\item  We confirm the presence of an anti-correlation between spin 
temperature and metallicity, using an absorber sample with measurements
of the covering factor. This is detected at $3.4\sigma$ significance 
for the sample containing both measurements and limits of \ts\ and 
metallicity (26 systems) and at $3.0 \sigma$ significance for a 
sub-sample (17 systems) with measurements of both \ts\ and metallicity.

\item  Although a wide range of spin temperatures are found at fixed
\nhi, the CNM fraction clearly depends on metallicity.  Absorbers
with low metallicity (for their \nhi) tend to have  high
spin temperatures and low CNM fractions.  Only the moderate-to-high
metallicity (above $\sim$ 1/10 solar) DLAs tend to show CNM fractions
comparable to, or in excess of, Galactic sightlines, at fixed neutral 
hydrogen column density.

\end{enumerate}

\section*{Acknowledgments} 

SLE is the recipient of an NSERC Discovery Grant which funded this
research.  NK and JXP have been supported by NSF grants AST-0709235
and AST-1109447. Support for HST program 12214 was provided by NASA
through a grant from the Space Telescope Science Institute, which is
operated by the Association of Universities for Research in Astronomy,
Inc., under NASA contract NAS 5-26555.  NK thanks the Department of
Science and Technology for support through a Ramanujan Fellowship. We
are grateful to Marc Rafelski for obtaining the HIRES spectrum of
J1431+3952, to Michael Palmer for providing the photometric redshift
of the nearby galaxy in the SDSS image and to Alain Smette for sharing
abundance measurements in advance of publication.  This work made use
of the Swinburne University of Technology software correlator,
developed as part of the Australian Major National Research Facilities
Programme and operated under licence.  The National Radio Astronomy
Observatory is a facility of the National Science Foundation operated
under cooperative agreement by Associated Universities, Inc. Part of
this work was done during a visit by NK to ESO; he thanks ESO for
support and hospitality.

\begin{appendix}

\section{Metal lines observed in optical echelle spectra}

\begin{figure*}
\centerline{\rotatebox{270}{\resizebox{12cm}{!}
{\includegraphics{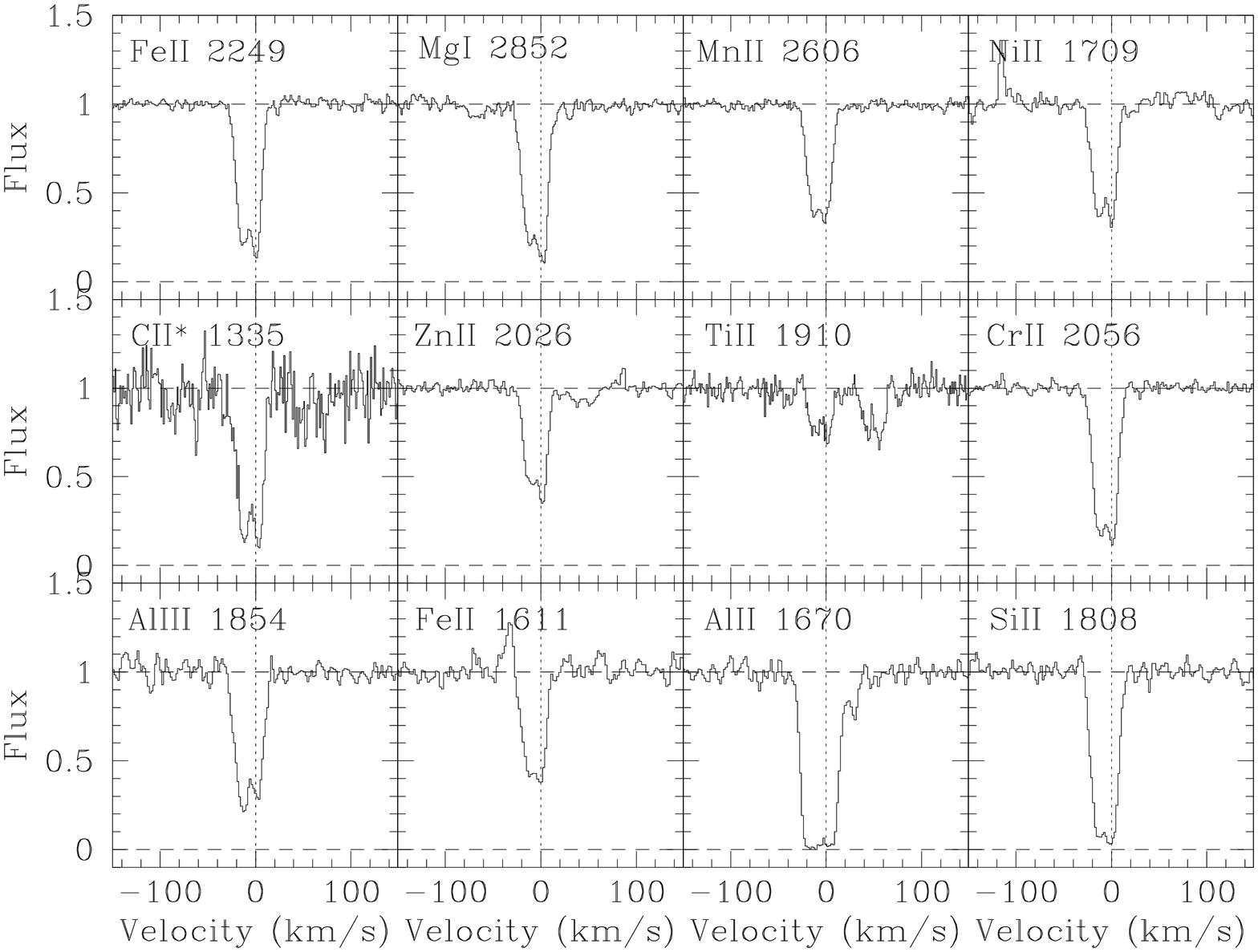}}}}
\caption{\label{UM305_metals}Metal line transitions in the DLA towards 
B0105$-$008.  The velocity scale is relative to $z=1.37104$. }
\end{figure*}

Examples of the echelle data obtained either from the
UVES archive or obtained by us from HIRES are presented
in Figures \ref{UM305_metals} to \ref{2355_metals}.

\begin{figure*}
\centerline{\rotatebox{270}{\resizebox{12cm}{!}
{\includegraphics{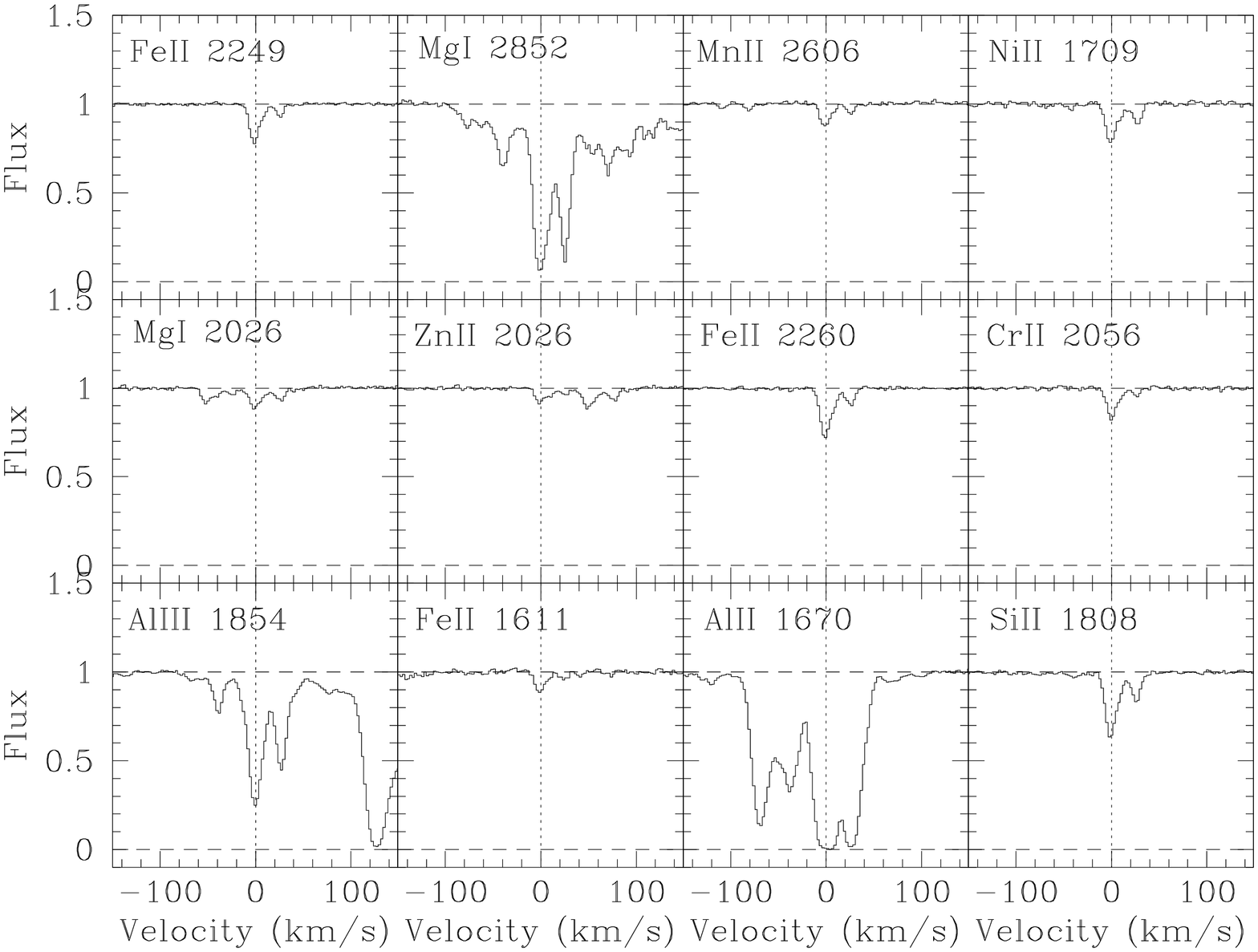}}}}
\caption{Metal line transitions in the DLA towards Q0237$-$233.  The
velocity scale is relative to $z=1.67235$. }
\end{figure*}

\begin{figure*}
\centerline{\rotatebox{270}{\resizebox{12cm}{!}
{\includegraphics{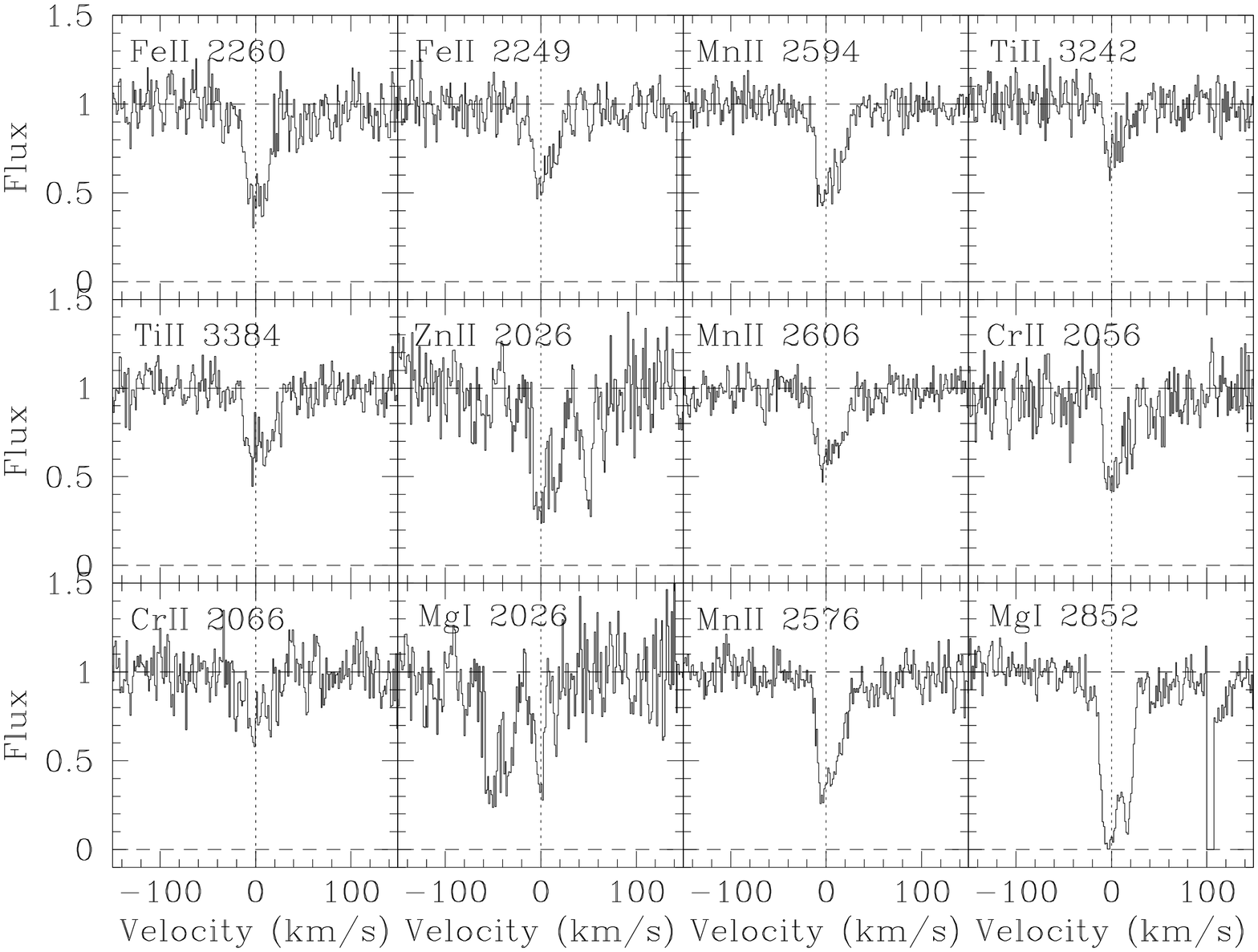}}}}
\caption{Metal line transitions in the DLA towards J1431$+$3952.  The
velocity scale is relative to $z=0.6018$. }
\end{figure*}


\begin{figure*}
\centerline{\rotatebox{270}{\resizebox{12cm}{!}
{\includegraphics{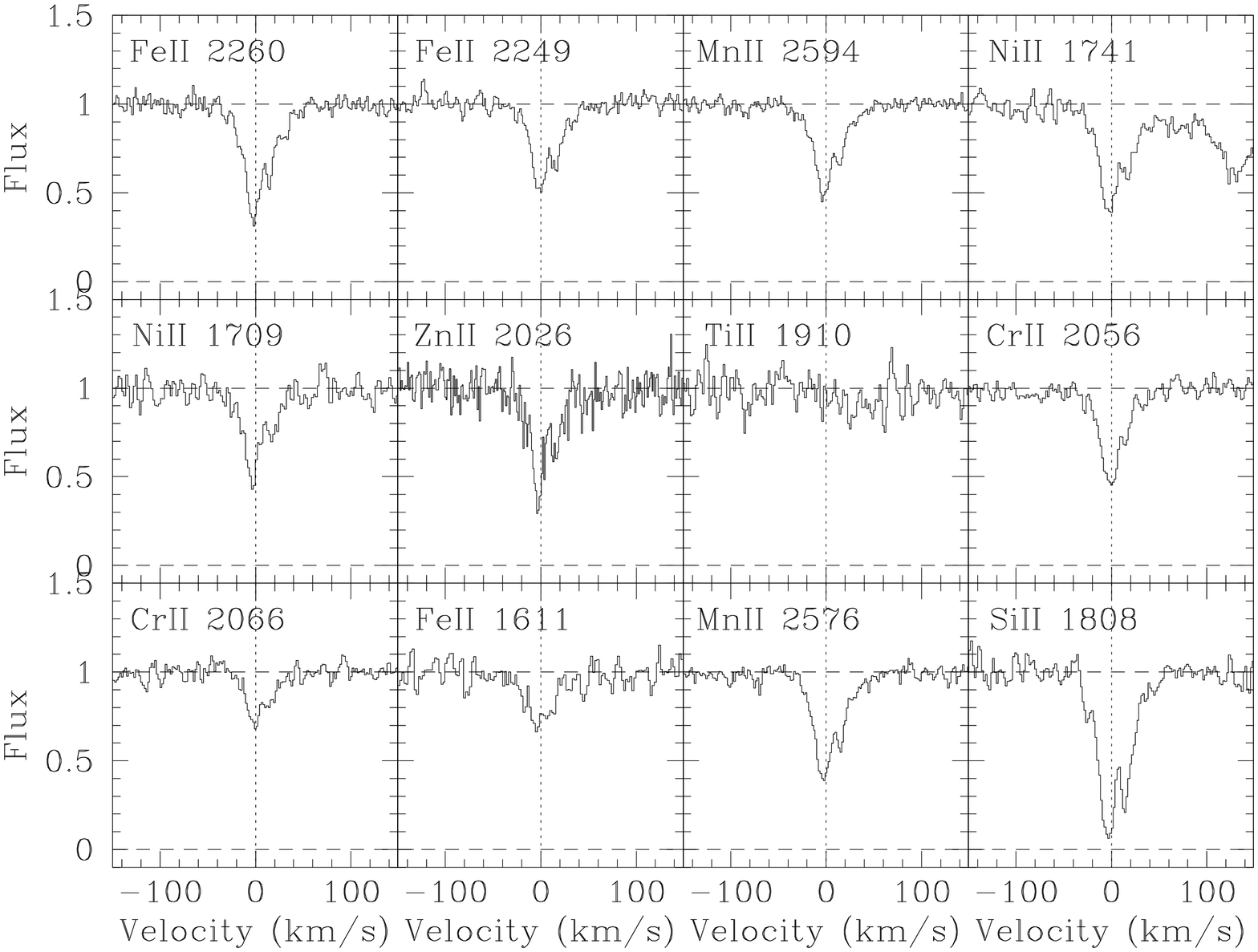}}}}
\caption{Metal line transitions in the DLA towards J1623+0718.  The
velocity scale is relative to $z=1.3357$. }
\end{figure*}

\begin{figure*}
\centerline{\rotatebox{270}{\resizebox{12cm}{!}
{\includegraphics{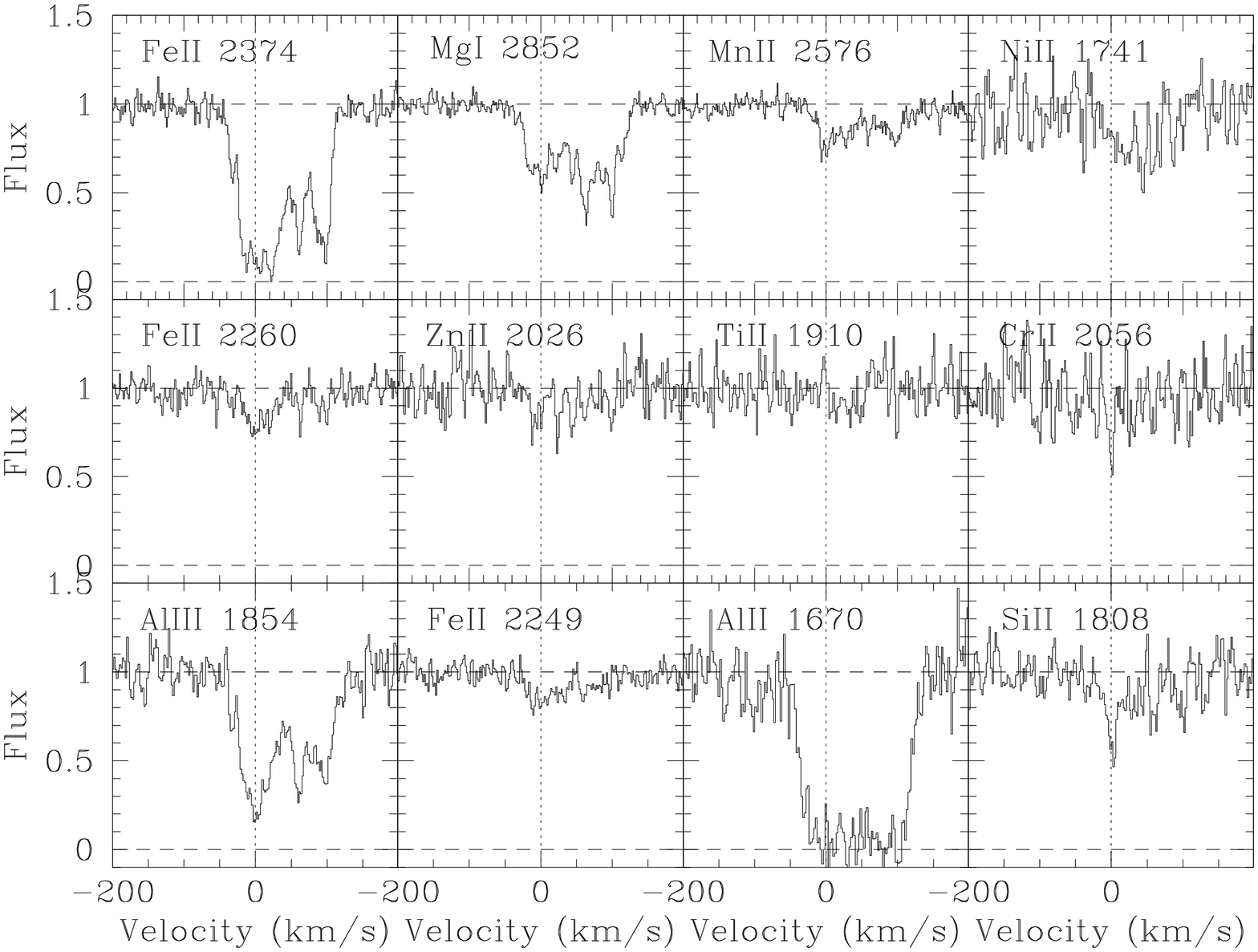}}}}
\caption{\label{2355_metals}Metal line transitions in the DLA towards
B2355$-$106.  The velocity scale is relative to $z=1.17230$. }
\end{figure*}

\end{appendix}

\end{document}